\documentclass[journal]{IEEEtran}
\usepackage{array}
\newcolumntype{L}[1]{>{\raggedright\let\newline\\\arraybackslash\hspace{0pt}}m{#1}}
\newcolumntype{C}[1]{>{\centering\let\newline\\\arraybackslash\hspace{0pt}}m{#1}}
\newcolumntype{R}[1]{>{\raggedleft\let\newline\\\arraybackslash\hspace{0pt}}m{#1}}
\usepackage{booktabs}

\ifCLASSINFOpdf
  \usepackage[pdftex]{graphicx}
  \DeclareGraphicsExtensions{.pdf,.jpeg,.png}
\else
\fi

\usepackage[dvipsnames]{xcolor}
\usepackage{multirow}
\usepackage{enumerate}

\begin{document}
%



\title{Authentication, Access Control, Privacy, Threats and Trust Management Towards Securing Fog Computing Environments: A Review}
%
%
%

\author{Abdullah Al-Noman Patwary,~\IEEEmembership{}
        Anmin Fu,~\IEEEmembership{Member,~IEEE,}
        Ranesh Kumar Naha,~\IEEEmembership{Member,~IEEE,}
        Sudheer Kumar Battula,~\IEEEmembership{ }
        Saurabh Garg,~\IEEEmembership{Member,~IEEE, }
        Md Anwarul Kaium Patwary,~\IEEEmembership{Member,~IEEE,}%
        ~and~Erfan Aghasian,~\IEEEmembership{Member,~IEEE.}%
\thanks{A. A. Patwary and A. Fu are with the School of Computer Science and Engineering, Nanjing University of Science and Technology, Nanjing 210094, China. e-mail: alnoman\_sub@hotmail.com, fuam@njust.edu.cn}
\thanks{R. K. Naha,  S. K. Battula, S. Garg, M. A. K. Patwary and E. Aghasian is with School of Technology, Environments and Design, University of Tasmania, Hobart, TAS, Australia. e-mail: raneshkumar.naha@utas.edu.au,  sudheerkumar.battula@utas.edu.au,  saurabh.garg@utas.edu.au, mdanwarulkaium.patwary@utas.edu.au, erfan.aghasian@utas.edu.au}
\thanks{Manuscript received XXX XX, XXXX; revised XXX XX, XXXX.}}

%
%

\markboth{Journal of \LaTeX\ Class Files,~Vol.~x, No.~x, x~x}%
{Shell \MakeLowercase{\textit{et al.}}: Bare Demo of IEEEtran.cls for IEEE Journals}
%



\maketitle

\begin{abstract}

Fog computing is an emerging computing paradigm that has come into consideration for the deployment of IoT applications amongst researchers and technology industries over the last few years. Fog is highly distributed and consists of a wide number of autonomous end devices, which contribute to the processing. However, the variety of devices offered across different users are not audited. Hence, the security of Fog devices is a major concern in the Fog computing environment. Furthermore, mitigating and preventing those security measures is a research issue. Therefore, to provide the necessary security for Fog devices, we need to understand what the security concerns are with regards to Fog. All aspects of Fog security, which have not been covered by other literature works needs to be identified and need to be aggregate all issues in Fog security. It needs to be noted that computation devices consist of many ordinary users, and are not managed by any central entity or managing body. Therefore, trust and privacy is also a key challenge to gain market adoption for Fog. To provide the required trust and privacy, we need to also focus on authentication, threats and access control mechanisms as well as techniques in Fog computing. In this paper, we perform a survey and propose a taxonomy, which presents an overview of existing security concerns in the context of the Fog computing paradigm. We discuss the Blockchain-based solutions towards a secure Fog computing environment and presented various research challenges and directions for future research.

\end{abstract}


%
\IEEEpeerreviewmaketitle

\section{Introduction}


%
%
%
%
\IEEEPARstart{T}{he} computational world has become very broad and complicated as our expectation is going beyond connecting people. We are about to approach a new era, where everything will be connected. With the swift development of technology, many individuals and organizations are starting to provide services to users with the help of their smart devices such as cell phones, home appliances, vehicles, wearable embedded devices, sensors, and actuators. The underlying work is performed by massive-scaled wireless sensor networks and realms of connected devices, which is aptly termed as the Internet of Things (IoT). IoT has achieved much attention over the last couple of years and has been enumerated as the predestination of the Internet. Technology consulting organization Gartner highlighted that the total number of connected devices by the year-end of 2020~\cite{gartner2017} would be more than 20 billion devices that exist across various consumers and business organizations. Moreover, Norton security organisation predicted that by 2025 there will be more than 21 billion devices \cite{symanovich2019future}. As IoT continues to flourish, a huge number of sensors have been devoted to diversified devices, which are swiftly leading to an increased amount of generated data and storage requirements on a regular basis~\cite{assunccao2015big}.

Although we are used to depending on the cloud for IoT application processing, the exponential growth of IoT devices continues to generate huge amounts of data, which means we will be unable to depend on any central entity such as the cloud computing paradigm to process these huge amounts of data. The Fog computing paradigm is evolving to serve various services while simultaneously managing numerous sensors, actuators, users, processes, and connectivity by placing processing facilities closer to users. Also, the edge devices generate data from their designated areas and link with each other or transmit to the neighboring Fog nodes for supplementary analytics and decisions. The Fog computing paradigm can solve the time-sensitive application processing limitations of the cloud as well as supporting IoT applications. Fog devices reside at the network edge to facilitate computing services near to the users and deliver services as well as applications for billions of connected devices. This helps to support real-time processing, storage and networking facilities at the edge level~\cite{bonomi2012Fog}. 


Since smart devices or Fog devices are categorized as resource constraints, the Fog computing paradigm will face many challenges such as the limitations of storage, bandwidth, battery, and computation power, which leads to obstruction in the rise of IoT. To overcome the encumbrance of these limitations, the cloud computing paradigm is perceived as a talented computing archetype, which can distribute services to the edge via the cloud in terms of Infrastructure as a Service (IaaS), Platform as a Service (PaaS) and Software as a Service (SaaS) solutions which offer applications and services with resilient resources at low costs~\cite{kapil2017cloud}. Over the last decennium, cloud computing has obtained an immense reputation among researchers. Real-time IoT application services and information access are possible any time and anywhere via this paradigm. Cloud computing also offers diverse features to users such as ease of access to information, cost efficiency, quick deployment, backup, and recovery. Although cloud computing has fulfilled most of the demands of modern technology, it may not be a suitable solution as there are still unresolved problems, whereas IoT devices and applications need to be processed swiftly. This is beyond the existing capabilities of cloud computing. Hence, security and privacy, data segregation, mobility support, low latency, location-awareness, geo-distribution, and real-time applications are required for IoT applications. While Fog computing offers a much more advantageous system as opposed to cloud-based systems, there are several security issues at hand which can cause interruptions to the way deployment is carried out using Fog computing. 

\begin{figure*}[htbp]
	\centering
	\includegraphics[width=6.5in]{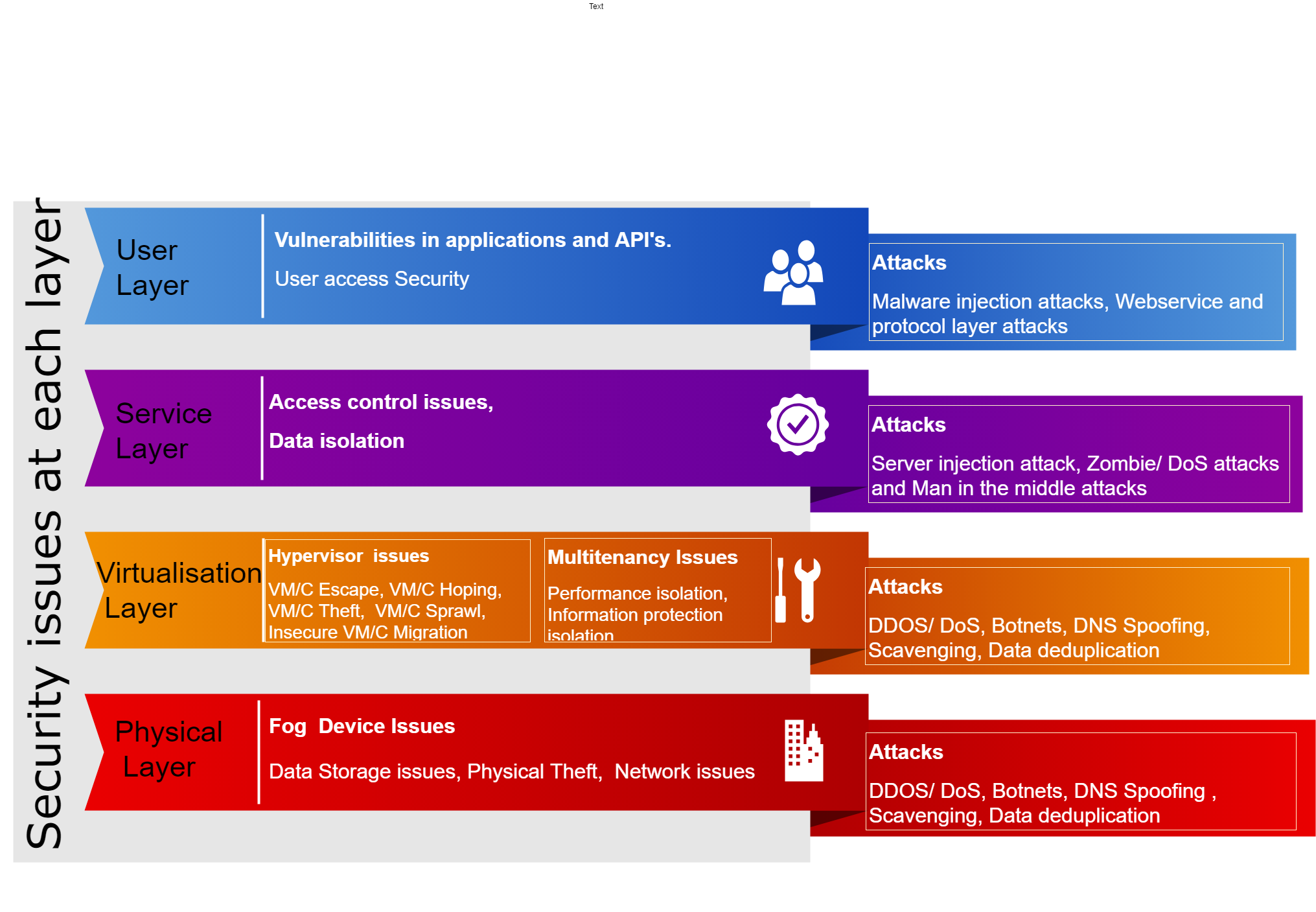}
	\caption{Security issues and attacks at each layer}
	\label{fig_Fogsecis}
\end{figure*}

In reality, due to the associated privacy and security risks for cloud-based systems, nearly 74\% of Information and communications technology (ICT) executive officers have rejected adopting cloud computing~\cite{zissis2012addressing}. Fog computing is not at a mature stage and continues to face new challenges due to its exclusive features. In the Fog computing environment, most devices are managed and maintained across different users. The Fog computing paradigm uses idle resources generated from user devices. These devices are not audited by any standard body, which raises security concerns in the Fog environment. On the other hand, secure and fast authentication mechanisms are required for Fog since many devices are involved in the Fog application processing. Furthermore, we have to be very concerned about access control since most of the application processing is carried out in the user devices. The security issues across various layers of the Fog computing environment is presented in Fig. \ref{fig_Fogsecis}. 


\subsection{Existing related surveys on Fog computing security}
There has been a variety of techniques proposed in the literature to address the security issues of the emerging Fog computing. Most of these research papers either presented Fog security concerns or merely focused on one aspect of Fog security. Here, we have summarized and given a concise overview with regards to Fog security by combining the opinions across several of these research works.
 

Yi et al.~\cite{yi2015security} briefly examined various security issues and tried to identify various challenge domains corresponding to the solutions of the Fog computing environment. Zhang et al.~\cite{zhang2018security} discussed and analyzed the adhering potential security and trust issues, and explored solutions which are currently available for those issues. Khan et al.~\cite{khan2017Fog} explored common security gaps in Fog computing from the existing surveys.  Alrawais et al.~\cite{alrawais2017Fog} investigated and discussed the various privacy and security issues in Fog computing environments. Rauf et al.~\cite{rauf2018security} discussed IoT, Fog and their security issues. Stojmenovic et al.~\cite{stojmenovic2014Fog} investigated intrusion detection and authentication techniques in Fog computing. Wang et al.~\cite{wang2015Fog} presented and discussed the concerns and challenges in Fog forensics and security. Recently, Roman et al.~\cite{roman2018mobile} explored potential threats associated with the mobile edge, mobile cloud, and Fog computing. 

In current literature, there is a gap in the aggregation of all Fog security-related issues. None of the literary works presented a critical evaluation of all aspects of Fog security, as has been done in this paper. Neither did they discuss Fog security issues from the auditing perspective. Different studies regarding Fog computing security and privacy did not cover the various security issues related to the Fog computing architecture and its environment. In this paper, we will explore and explain various security concerns related to the Fog computing environment from the auditing perspective. Since Fog computing extends to the cloud system, therefore, most of the cloud computing security concerns~\cite{takabi2010security} are being inherited and impacts Fog computing as well. We have focused our attention on significant security, threats and attack issues such as trust management, privacy, authentication, and access control. We linked these security concerns with Fog and explained how these concerns could affect Fog security. In addition, we discussed how blockchain could mitigate some Fog related security issues. We have systematically focused our attention on significant security and threat-attack issues from several selected sets of papers to provide a detailed landscape in this field./


%


\subsection{Research Motivation} Cloud computing is already recognized by its widespread deployment amongst its targeted environment. However, it faces numerous obstacles such as latency, bandwidth, Quality of Service (QoS), trust, security, privacy, trust, threats, and attacks, etc. during the early stages of its deployment. Therefore, privacy and security are the key challenges for the cloud computing paradigm. In the case of Fog computing, it was inaugurated as a new computing paradigm, which has emerged over the last few years as a bridge between cloud data centers and edge devices or IoT devices. The main aim of Fog computing is to improve the existing problems of cloud computing by improving the communication latency, real-time processing, privacy and security. Nevertheless, Fog computing also faces many privacy and security concerns as it is in its early stages. User devices and end devices are the main components for computation in the Fog environment, which is not usually audited by any security standard. Therefore, the key aim of this work is to come up with a methodical review on state-of-the-art approaches and techniques in accosting Fog computing security and privacy issues from the auditing perspective and pinpoint challenges as well as the possible direction for researchers and application developers.

\subsubsection{Paper Selection Approaches} 
\begin{figure*}[!t]
	\centering
	\includegraphics[width=5.3in]{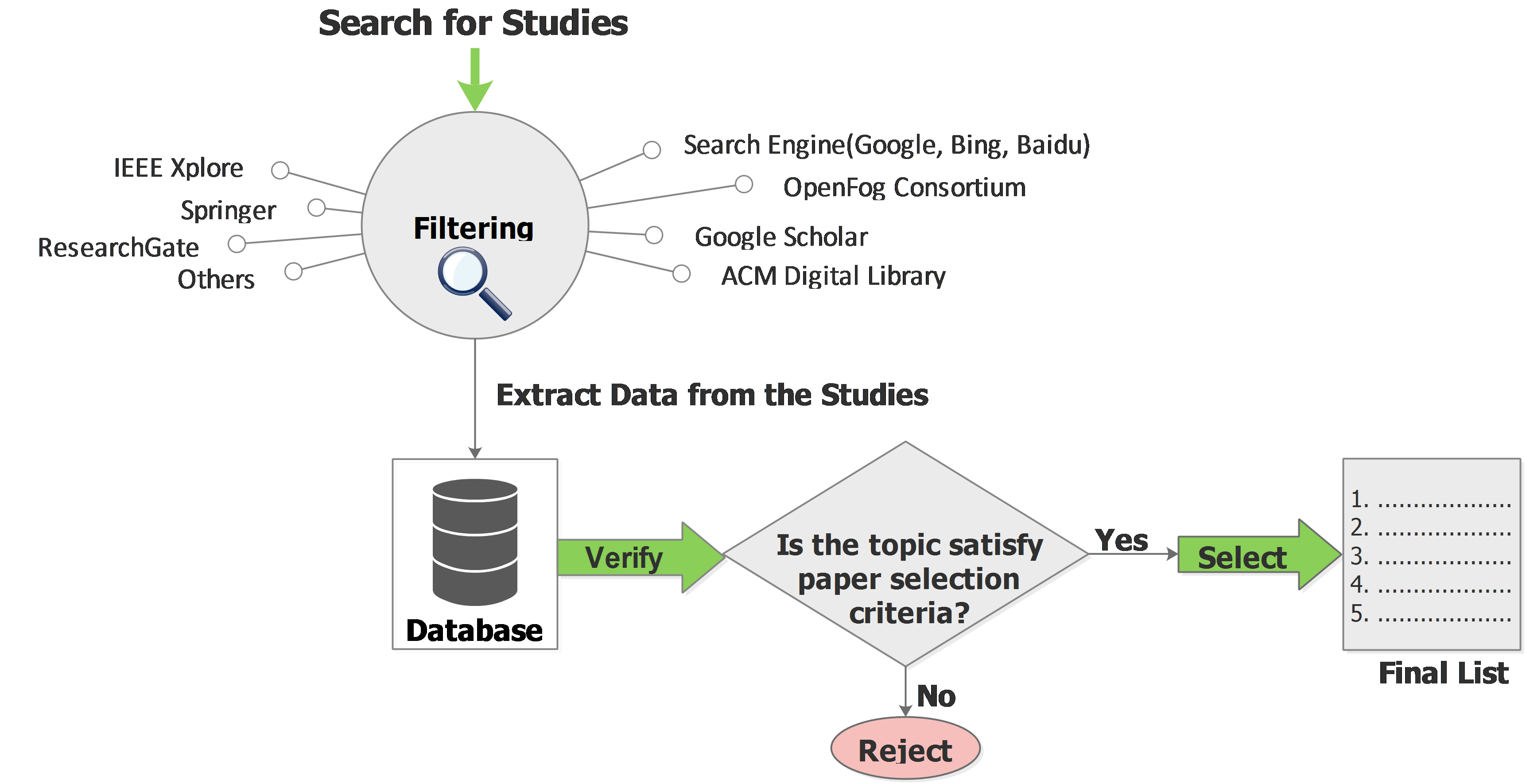}
	\caption{Paper selection process.}
	\label{fig_papsel}
\end{figure*}

To exploit the coverage of the searched literature in this work, we began by identifying the most used alternative words and synonyms in the research questionnaire. Therefore, we conducted our selection strategy based on our proposed taxonomy and Table  \ref{paper_sel_proc} searching criteria. We first categorized the current research security issues and challenges for Fog computing into six categories: 1. Trust, 2. Privacy, 3. Authentication, 4. Access-Control, 5. Threats and Attacks, 6. Security Audit. We also looked into the security issues and solutions of other areas such as cloud, edge computing, and blockchain which could suit the Fog computing environment. In order to focus on the most relevant articles based on the aims of our research, we also constructed different search strings using Boolean AND and OR operators. Then, we conducted a manual search (Fog computing security issue or privacy and security issue in Fog computing), using different search engines such as Google, Bing, Baidu etc. in the area of cloud, Fog computing security based on the search criteria in Table  \ref{paper_sel_proc}. The same approach was applied in renowned scientific research databases such as Google Scholar, ACM Digital Library, IEEE Xplore, Springer, Science Direct and ResearchGate. Fig. \ref{fig_papsel} presents our paper selection approach. We used the tool Mendeley and Google Scholar to manage citations from all extracted articles. We conducted our paper selection and evaluation based on the various criteria as shown in Table \ref{paper_sel_proc}. 

\begin{table}[htbp]
	\centering
    \small
	\caption{Paper Selection Criteria}
	\label{paper_sel_proc}
	\begin{tabular}{L{1cm}|L{6cm}} 
		\toprule
		\textbf{Sl. No.} & \textbf{Criteria}\\ \hline
        01	& Relevant to study of the cloud or Fog computing \\ \hline
        02	& Directly or indirectly related to cloud and Fog computing security \\ \hline
        03	& Fog computing security issues \\ \hline
        04	& Security and privacy issues in Fog computing \\ \hline
        05	& Security and trust issues in Fog computing \\ \hline
        06	& Authentication and authorization in Fog Computing \\ \hline
        07	& Authentication and access Control in Fog Computing \\ \hline
        08	& Privacy preservation in Fog computing \\ \hline
        09	& Threats and attacks issues in Fog computing \\ \hline
        10	& Security auditing standards in Fog computing\\ \bottomrule
    \end{tabular}
\end{table}

\subsubsection{Evaluation of Results} After the initial exploration using several search strings from the sources above, we found almost 220 relevant papers and articles. After searching, filtering, inclusion and exclusion reviews, 127 articles were matched from the first filtration. With respect to our taxonomy, we have separated all these papers into various partitions.


\subsubsection{Research Questions}

This work is going to answer the following research questions:

\begin{enumerate}[Q1]
    \item What are the different security issues in Fog which need further investigation?
    \item What are the all security aspects of Fog and how to categorize them?
    \item How current research works addressed Fog security concerns? What are the other possible solution and what security concerns need attention to the research community?
\end{enumerate}

Section II to IV are answering our first two research question. The third research question is answered by section IV, V and VI.


\subsection{Our Contributions}
This survey is intended to provide an exhaustive review across current studies by covering all related Fog security issues and challenges. This work also concentrates on constructing a review of Fog computing with a focus on the related challenges and security issues from the auditing perspective. The principal contributions of this study can be recapped as follows:
\begin{itemize}
\item Propose a taxonomy based on various security issues such as authentication, access control, privacy preservation, trust management, threats, attacks, and security auditing which are challenging for the Fog environment.
\item Highlight and discusses various threats and attacks which might be severe in the Fog environment.
\item Discuss probable challenges and future research directions in Fog computing with respect to security.
\item Explain how blockchain and auditing could help to mitigate Fog security challenges.
\end{itemize}

The rest of the paper is organized in the following manner - Section II provides an overview of Fog computing. Section III discusses the network and data security issues. Section IV demonstrates the proposed taxonomy on security issues in Fog computing. Section V discussed blockchain technologies in Fog and present how blockchain technology can be utilized to improve Fog security. Section VI and VII present the research challenges, future research directions, and conclusions.

\section{An Overview of Fog Computing} 
\begin{figure}[!t]
	\centering
	\includegraphics[width=3.5in]{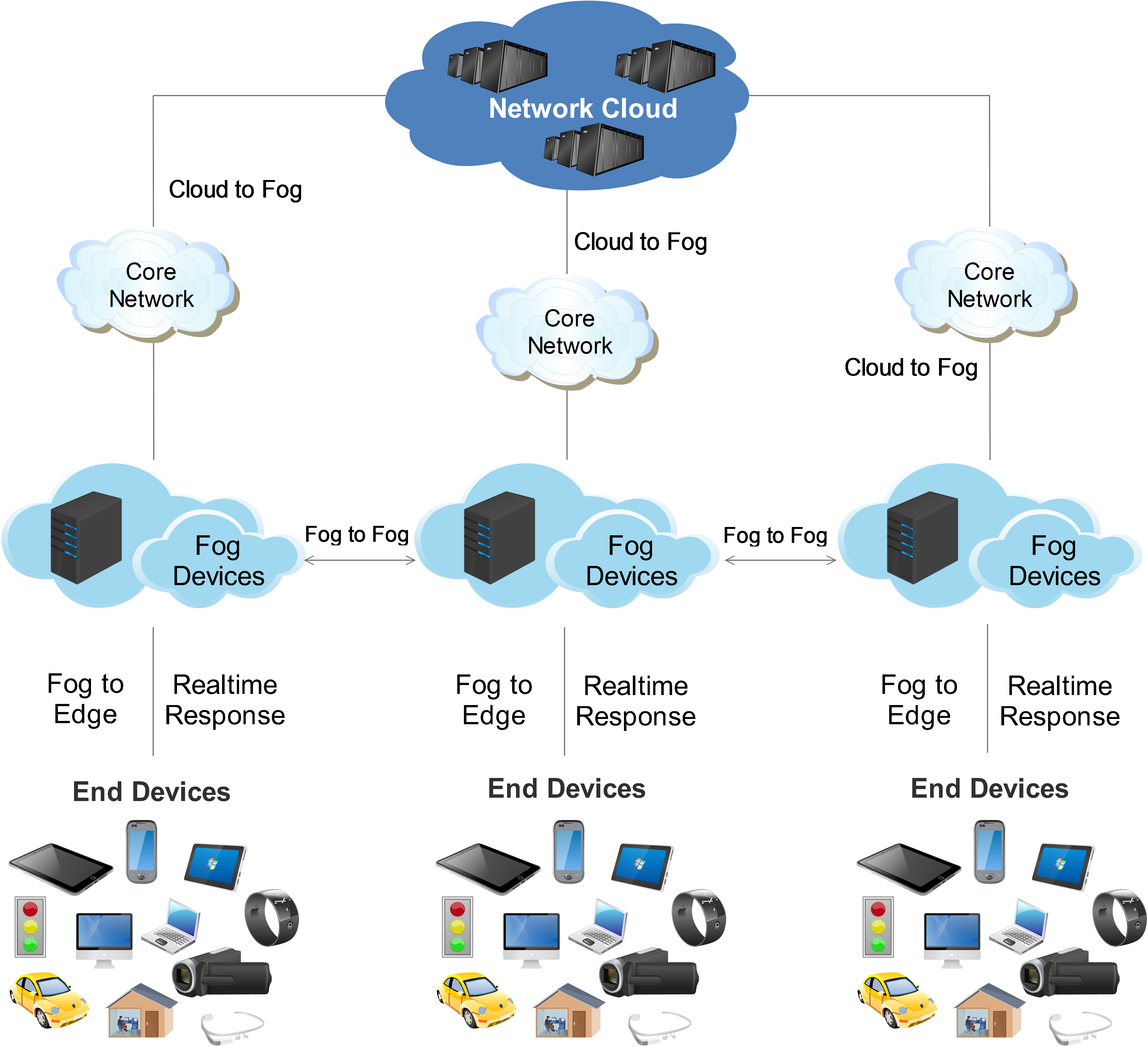}
	\caption{The Architecture of Fog Computing.}
	\label{fig_Fogco}
\end{figure}

Fog computing ideally demonstrates the concept of a distributed network environment that connects two different environments and is closely linked with cloud computing and IoT. This new computing paradigm was initially and formally introduced by Cisco to extend the cloud network to the edge of the enterprise network~\cite{bonomi2012Fog}. The architecture of a Fog environment has three layers - the IoT layer, the Fog layer and the Cloud layer, as shown in Fig.~\ref{fig_Fogco}. The IoT layer consists of a massive amount of sensors and end devices. This layer is liable for collecting and sending the data generated from devices to the Fog devices in the Fog layer. The Fog devices in this layer then process the received data and send the results to the cloud to store for future use. Individuals or organizations are providing Fog devices to process the applications in a Fog environment by contributing their idle resources. The providers should compensate for their offered resources based on the usage in a way that both provider and user will be benefited \cite{battula2019micro}. 


In literature, there exist similar Fog like technologies such as Edge Computing, Mobile Cloud Computing (MCC), Cloud Computing, Mobile Edge Computing (MEC), Cloudlet, Fog Dew Computing, Dew Computing and Micro Data Centre  \cite{naha2018Fog,khan2017Fog}. However, the key difference is that it creates an enormously virtualized platform that offers diversified computation, storage, and network services to its clients via unused end-device resources. With the features and characteristics of the Fog computing continuing to improve, the performances of a wide range of domains across different real-time IoT specific applications such as City: smart office, smart home, smart waste management; Electricity: smart grid; smart metering, Health: smart health care system, Transportation: smart vehicle accident prevention; traffic flow maintenance; Smart Traffic Light System (STLS); Traffic control system, Entertainment: real-time video streaming and gaming systems are shown in Table \ref{app}.
\begin{table}[]
	\caption{Examples of Fog application}
	\label{app}
	\begin{tabular}{L{1.5cm}|L{1.5cm}|L{4cm}} 
	\toprule
\textbf{Application domain}&\textbf{Application service}&\textbf{Description} \\ \hline
Smart city~\cite{soliman2013smart}~\cite{zanella2014internet}&Smart home and Smart office&Provides automation control in home to control the electrical appliances and security and alarm systems\\ \hline
Electricity~\cite{kyriazis2013sustainable}\cite{ejaz2017efficient}&Smart grid and Smart metering&Provides monitoring and tracking service of energy hourly or day wise etc. \\ \hline
Healthcare~\cite{yuehong2016internet}&Smart health monitoring&Provides continuous monitoring of glucose, blood pressure, pulse rate  etc. \\ \hline
\multirow{2}{*}{Entertainment 
}&Augment Reality&Provides the best user experience in Augmented  Reality \\ \cline{2-3}
~\cite{xu2014non} &Real-time video streaming and gaming  system 
Entertainment & Provides the best user experience in video streaming and gaming systems \\ \hline
\multirow{4}{*}{Transportation 
}&Smart vehicle & Driverless vehicles \\ \cline{2-3}
&Smart navigation & Suggests best routes and dynamic rerouting \\ \cline{2-3}
 ~\cite{tammishetty2017iot} ~\cite{gerla2014internet}&Road condition detection & Auto detects the condition of roads and adjusts the parameters to drive according to it \\ \cline{2-3}
&Smart traffic lights&Reduce the traffic jams across the junctions \\ \bottomrule
\end{tabular}
\end{table}

The features and characteristics of Fog computing are as follows~\cite{bonomi2012Fog}: 
\begin{itemize}
\item Support Geographic Distribution
\item Location Awareness
\item Low Latency
\item Heterogeneity
\item Decentralization
\item Large Scale QoS-aware IoT Application Support
\item Mobility Support
\item Interplay with Cloud
\item Context Awareness
\item Online Analytics
\item Predominance of Wireless Access
\item Close to the end users
\item Save storage space
\item Higher Scalability
\item Save Bandwidth
\item Real-time Interaction
\item Data security and privacy protection
\item Low energy consumption
\end{itemize}


However, Fog computing has provided numerous other issues and challenges such as security and privacy. The technical distinctions between Fog and cloud computing from a security aspect are exhibited in Table~\ref{cloudFogdiff}. The OpenFog Consortium, technology giants, researchers and developers are strongly trying to mitigate these issues. Therefore, if they were able to attenuate all these issues, then it would be deemed capable to deal with the constantly increasing number of networked computational devices. This would then make the Fog platform the future of computing.
  
In accordance with the study of Fog computing characteristics, we have illustrated a differential table based on cloud and Fog features - Table \ref{cloudFogdiff}. Finally, we have pointed out a few challenges that exist for the current cloud technology. Therefore, we have also illustrated a table and highlighted how Fog eliminates these challenges - Table \ref{cloudFogcha}.

\begin{table}[htbp]
	\centering
    \small
	\caption{Technical difference between Fog and cloud in a security perspective}
	\label{cloudFogdiff}
	\begin{tabular}{L{2.5cm}|L{2cm}|L{2.5cm}} 
		\toprule
		\textbf{Attributes} & \textbf{Cloud Computing} & \textbf{Fog Computing}\\ \hline
        Security management	& Centralized	& Distributed\\ \hline
        Security concerns & General servers & Heterogeneous devices \\ \hline
        Attack and threat level &	Low	& High\\ \hline
        Security domain &	Within the Internet	& At the edge of the local network\\ \hline
        Security pattern &	No user defined security & User defined security\\ \hline
        Security Audit and Analysis & Static or manual approach  & Software based automated dynamic and real-time approach\\
        
        \bottomrule
    \end{tabular}
\end{table}

\begin{table}[htbp]
	\centering
    \small
	\caption{Security challenges at Fog}
	\label{cloudFogcha}
	\begin{tabular}{L{2.5cm}|L{5.5cm}} 
		\hline
		\textbf{Challenges} & \textbf{Role of Fog}\\ \hline
        Security of computing and access control & With Fog, the computation, process, storage and control of sensitive tasks are done as near as possible to the end user's device. In this distributed environment, all threats and attacks first need to be faced as Fog nodes, where Fog nodes are able to identify all illegitimate activity and can prevent any incidents before they are passed through to the system. \\ \hline
        
        Security of data storage and users privacy & In Fog environments, data is originated from, or to sent to the end-user devices which are managed and preserved via secure Fog nodes. Hence, the data would be better preserved than stored in the user's device and more available than if it was maintained in remote data centers. \\ \hline
        
        Security of communication and networking system & A Fog network is connected by an immense collection of Fog nodes, and it can provide uninterrupted secure communication and networking services by residing near the end user's device. Fog reduces the chances of various network and communication attacks. \\ \hline
        
        Security of the resource-constrained IoT devices & A lot of IoT devices or end devices has limited resources. Hence, due to these limited resources, the IoT devices have little or no capability to defend themselves from sophisticated cyber-attacks. Fog nodes and cloud servers together can provide multi-level protection, i.e. "defense-in-depth". \\ \hline
      
        Real-time incident response services & In Fog networks, the Fog nodes are able to provide real-time incident response services that notify the IoT system without disruption of any services. \\ \hline

       Security challenges in the edge network & Because of the lack of available resources to end devices, Fog can manage and update security mechanisms such as authentication, access control, trust management, etc. Therefore it can also protect devices that cannot protect themselves adequately.\\ \hline
        
        Security credentials and software up to date & It is impractical to require that all the devices are connected several times a day to cloud for the security credentials and software to be updated. However, Fog nodes are able to manage security credentials and software updates on a large number of devices simultaneously, based on their criteria without downtime. \\ \hline
        
        Monitor the security status & In the IoT environment, it is crucial to be able to notice trustworthy processes, whether the devices and systems are operating safely and securely. Many of today's hackers send false status messages that make operations appear normal. Fog provides a scheme to monitor security status in a trustworthy manner and can detect these types of attacks.	\\ \hline

    \end{tabular}
\end{table}

%
%
As Fog devices are much more distributed and belongs to different users, security auditing is very important. In order to audit the security of Fog devices, we need to explore the network and data security issues related to  Fog.  

\section{Fog Network and Data Security}
Ensuring security for both network and data in Fog is a challenging task due to the vastly distributed nature of Fog computing. Most of the Fog devices are wireless, and data is processing in the user's devices. This section discussed the network and data security of Fog in detail.

\subsection{Network Security}
Due to the massive deployment of wireless networks in the Fog environment, ensuring security in these networks is a mandatory concern. Wireless networks are prone to attacks such as jamming, sniffers, spoofing, Man-in-the-middle (MITM), etc. These attacks can affect the wireless network security of Fog computing, which can take place between the cloud to things continuum. In general, the users trust the network configurations and data generated by the network traffic which is usually managed manually by a network administrator~\cite{tsugawa2014cloud}. As Fog nodes placed at the edge of the network, therefore, it would be an unmanageable task for the network administrator. In such a scenario, the Software Defined Network (SDN) will increase the scalability of the network and decrease the cost. Hence, SDN would be a preferable solution in Fog computing~\cite{yi2015security}. In Fog computing, SDN can provide features for network security, for example monitoring networks and Intrusion Detection System (IDS), as well as watching the traffic routes which is referred to as CloudWatch~\cite{shin2012cloudwatcher} and OpenFlow~\cite{mckeown2008openflow}. It also helps to isolate the traffic and manage prioritization to prevent attacks from network resource access controls and congested networks.  Klaedtke et al.~\cite{klaedtke2014access} proposed a method for access control that was based on OpenFlow and for a network resource sharing system. The authors~\cite{press2014idc}, proposed an OpenWifi, which gave authentication to the guest users by letting them have access to the Fog node router in context with the security issues.

\subsection{Data Security}
In Fog computing, data generated by IoT or edge devices are gradually increasing respectively with the number of IoT devices. Due to a lack of adequate resources for IoT devices, it is hard to process all the data on IoT devices~\cite{alrawais2017Fog}. IoT devices send the generated data to the nearby Fog node. After that, this node divides the generated data into several segments and forwards them to multiple Fog nodes for further processing. During this division and distribution time, the data could be altered or manipulated by attackers. Therefore,  the integrity of the data must be ensured. Hence, the encryption and decryption process is not easy to implement due to associated resource constraints. In this case, light-weight encryption and decryption techniques would be a compatible solution~\cite{lee2015security}. However, user data is being outsourced as well the user's data control which is handed over to the Fog node. This still brings about the same security threats associated with cloud computing. In this circumstance, there might be a chance to lose or modify the outsourced data. In addition, illegitimate third parties with malicious interests might misuse the stored data. To mitigate these threats, a proposed solution is to present auditable data storage services, which are applicable for cloud computing data protection. In the context of a cloud storage system, a well-known technique is a homomorphic encryption and searchable encryption, which could be used to accumulate and ensure integrity, confidentiality and verifiability to permit a client to investigate the data which is stored on untrusted servers ~\cite{lu2012eppa}. Yang et al.~\cite{yang2012data} surveyed the existing research work related to auditing data storage services in the context of cloud computing. Eventually, from the circumstances above, there is still no proposed method that can meet the criteria based on a three-tier architecture for Fog computing. Nonetheless, it is a challenging task to design a secure storage system, which will satisfy all requirements (dynamic processing, low-latency, high-scalability, etc.) and support smooth communication between the Fog and cloud environments. To detect network and data attacks in Fog we need to employ an Intrusion Detection System (IDS) across various layers.
 
Intrusion Detection System (IDS) is extensively used in cloud systems to identify and help protect from attacks, such as Denial of Service (DoS) attacks, insider attacks, port scanning attacks, flooding attacks on the VM (Virtual Machine), man-in-the-middle (MITM) attacks, hypervisors, as well as numerous systems ~\cite{modi2013survey}. It can be deployed under Supervisory control and data acquisition (SCADA)~\cite{maglaras2016combining}, cloud~\cite{modi2013survey}, smart grid system~\cite{valenzuela2013real}~\cite{qin2013defending} etc. It can also monitor, detect intrusive behavior of possible attackers, as well as analyze log files, access control (AC) policies, and user access credentials. In three-tier architecture Fog computing, IDS must be deployed in the cloud, Fog, edge for monitoring, analysis of traffic and intrusive activities of cloud servers, Fog nodes and edge devices. However, establishing security alone is not enough to provide the necessary protection against the propagation of viruses or malware from vulnerable nodes to other parts of the system. With regard to this situation, there may arise challenges such as corrective responses, alarm parallelization, false alarm controls, and real-time notification~\cite{anwar2017intrusion}. A probable solution could be to deploy a perimeter IDS that coordinates different IDS in the Fog system~\cite{cruz2016cybersecurity}. On the contrary, while ensuring security in the Fog computing environment through IDS, several challenges may arise in terms of providing low-latency requirements~\cite{yi2015security}.

\subsection{Security Standards in Fog}
Security standards form a vital part in maintaining protection for information systems. These standards are responsible to define the scope and security functions and features needed, as well as policies, in order to manage the information and human assets. Standards also help to evaluate the effectiveness of security measures and maintain the criteria for ongoing assessments of security. It is a necessity to consider proper security standards and commonly used security practices in the Fog computing environment in order to develop a feasible choice for the enterprise community.


IEEE 1934~\cite{ieee2018ieee} is a standard reference architecture for Fog to satisfy data-intensive application requirements. This architecture was proposed based on eight key attributes of the system, for example, RAS (reliability, availability, and serviceability), scalability, autonomy, openness, security, agility, hierarchy and programmability. For auditing purposes, we need to figure out the taxonomy of Fog security issues. By which, we can then identify what to audit and how to perform auditing in Fog by following recommended standards.

\section{Taxonomy of Security Issues in Fog Computing} 
\begin{figure*}[!t]
	\centering
	\includegraphics[width=6.5in]{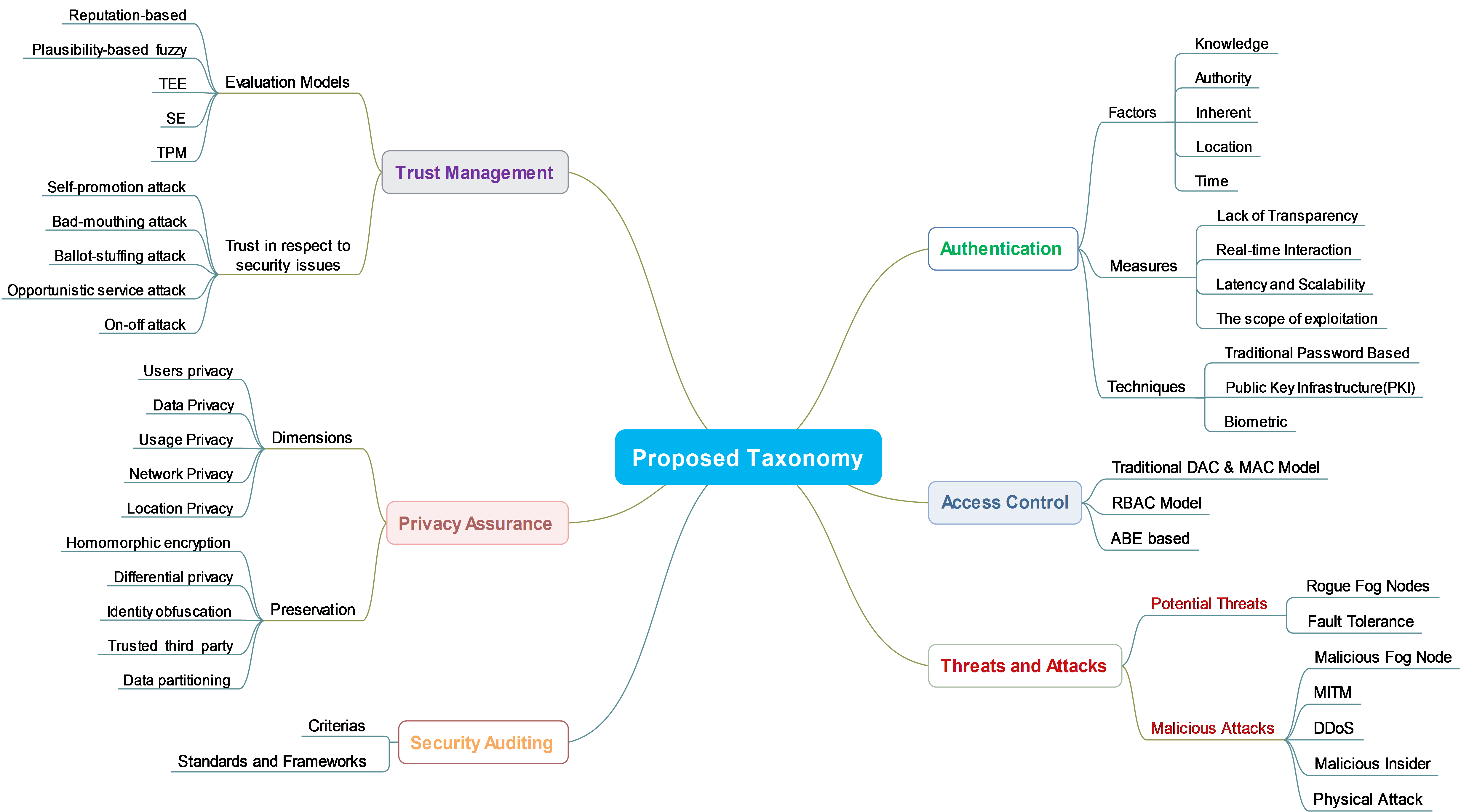}
	\caption{A Taxonomy of Security Issues of Fog Computing.}
	\label{fig_tax}
\end{figure*}

Fog is an augmentation of cloud computing which has many security issues. In this study, we have proposed a taxonomy, which is based on various security issues such as trust management, privacy assurance, authentication, access control, threats, attacks and vulnerabilities adhering to the Fog computing environment for auditing purposes. In the trust management section, we have discussed trust, the scope of trust, trust model and the potential attack on the trust computation area. In the privacy assurance section, we have discussed different privacy issues and privacy preservation techniques. In authentication, our observation relates to authentication domains, methods and potential attacks on the authentication processes. In the access control section, we identified the controlling area, requirements and access control methods. Finally, we summarized several threats, attacks, and vulnerabilities. This taxonomy offers a better understanding of Fog security issues to the research community and enterprises. Fig.~\ref{fig_tax} represents the proposed taxonomy and concise derivation of each section in the taxonomy, which will be described in the following subsections.

\subsection{Trust and Trust Management in Fog Computing}
The definition of trust does vary across different fields. Trust is the level of undertaking that an entity will treat in an appeasing way~\cite{li2007trust}. Although this definition does not represent the proper trust definition according to the field of computing, it can be characterized as an ``expectation that a device or system will faithfully behave in a particular manner to fulfill its intended purpose''~\cite{rahman2018find}. Therefore, trust can support the devices that failed to communicate with each other and desire to establish a new connection. A Fog node might be considered safe or unsafe by relying on their trust level.

Trust management is considered in order to establish trust between entities. It is a system or mechanism that takes place between two nodes in a network to established trust. It was first introduced by Blaze et al.~\cite{blaze1996decentralized}. They defined the problem of trust management as \textit{``the problem of figuring based on formulated security policies and security credentials if a set of security credentials of an entity satisfies the security policies''}. Trust management examines the way of collecting and storing information to ensure the trustworthiness of an entity. It can be measured with creation, updating or revoking the trust~\cite{cho2011survey}.

In Fog computing, the devices are responsible to provide reliable and secured services for end-users. In this case, there must have a definite level of trust between all the devices in the Fog network. Authentication plays an important part in forming a primary set of relations between the end user's device and Fog devices in the system. As devices can always breakdown or become vulnerable to malicious attacks. Authentication alone is not adequate to fix these problems. Fog computing has an aim to elevate the trustworthiness of the overall network. In cloud platform technology, the data centers are typically owned and maintained by cloud service providers. However, in the Fog platform, dissimilar parties may act as service provides as diverse deployment options exist in such systems~\cite{yi2015security} such as Internet service providers, Cloud services providers, and End-users. This flexibility makes obscure the required trust for Fog computing. Therefore, based on these circumstances, numerous problems arise in the Fog computing environment as follows:

\begin{itemize}
\item In the Fog environment client is a node that can apply the required services as presented by the Fog device. Hence, Fog devices are retained and upheld autonomously and operated by various organizations or parties. In such a case, Fog clients are required to be more vigilant in the time of communication with Fog nodes. Generally, different possessors preserve security in different ways, and the security amongst Fog devices positioned in the same organization may also be dissimilar in context. Therefore, from a Fog client's observation, Fog nodes indicate a potentially great threat.
\end{itemize}

\begin{itemize}
\item From a Fog node's perspective, the client is also considered as a potential threat. These services can comprise of various scripts or harmful cipher with destructive consequences to the Fog node's software or hardware.
\end{itemize}

\begin{itemize}
\item Data is collected from the Fog clients through the Fog network and it can be used for further work. However, after the data is collected from Fog clients, it might be corrupted or lost during the propagation process.
\end{itemize}

\begin{itemize}
\item Fog nodes can be deployed by anyone or any organization. Therefore, setting up a Fog node that may become a threat to the whole network may be complicated~\cite{stojmenovic2014Fog}. A rouge Fog device can send illegal data and run over the entire network, which can have undesirable influences on the entire network performance and amplify the packet loss. This compromises Fog nodes or rouge nodes which can hamper the legitimate nodes in the Fog network.
\end{itemize}

\begin{itemize}
\item Usually, Fog nodes can be installed or deployed near the end-users, so that Fog nodes are easily accessible and can be tampered with spontaneously. If node hardware or software is tampered with, it will become a potential threat for the entire network. Therefore, data that is shared with the tampered Fog device can be exposed or revealed to unauthorized entities.
\end{itemize}

\begin{itemize}
\item Any Fog device which is compromised can be a source from which originates malicious objects which can impact the reliability of the whole Fog network.
\end{itemize}

In such scenarios, trust helps to maintain the relations built upon preceding interactions of devices or entities. Trust must play a two-way responsibility in the Fog environment~\cite{mukherjee2017security}. First, the nodes that provide services to edge devices must be competent to authenticate the service requests to comprehend if the request is fake or genuine. Second, the edge devices that send or request data must be competent to authenticate the intentions of the node to guarantee its security. Therefore, applying the trust mechanism in the Fog environment permits Fog nodes, resource-limited IoT devices, and other Fog clients to identify the future behavior of one another. When identification of future behavior becomes probable, then Fog clients can easily choose a trusted Fog node that will provide the best services. As a sign of the problems presented in the solution of trust management, for a Fog system, there is a need to identify and detect all accidental or intentional behavior which can enable authorities to take the necessary action and rebuild the trust formation instantaneously~\cite{yi2015security}. The key factors that influence Fog computing are trust scope, trust characteristics and trust evaluation models.

\textbf{Trust Scope:}
Guo et al.~\cite{guo2017survey} demonstrated current methods of trust computation in the IoT system. They categorized the trust computing scheme into five scopes: aggregating, formation, update, propagation and trust composition. We can consider this scope of trust for the Fog computing environment as well. This segment will demonstrate each of these scopes in detail as below:

\begin{itemize}
\item \textbf{Trust Aggregation:} collect all the recommendations from others and combine them with one's own experiences in the trust computation which might be essential. Trust Aggregation elects how this is accomplished.
\item \textbf{Trust Formation:} this defines the way to enable a combination of trust properties by trust composition. Some methods just study one property, and others reflect a mixture of some properties.
\item \textbf{Trust Update:} it shows how often the trust values are updated. Periodical updates and Event-driven are two key methods.
\item \textbf{Trust Propagation:} it decides on how to select a distributed or centralized process to compute and store the trust.
\item \textbf{Trust Composition:} it defines a group of trust properties. It chooses what components have been used in the trust computation process. Social trust and service quality are the two key elements.
\end{itemize}

\textbf{Characteristics of Trust in Fog Computing:}
This section describes the characteristics of trust in Fog computing. Various characteristics of trust that help develop trust relationships related to the understanding of Fog computing much further. The authors~\cite{pranata2012holistic} defined a few characteristics, which can be retained for the Fog environment.

\begin{itemize}
\item \textbf{Is trust dynamic? } Trust requires to be dynamic because of two reasons. First, the Fog system network topology is changing continuously as new devices join or leave concurrently on the Fog network. Then, devices in the network may deflect their behavior successively. Therefore, trust should be monitored uninterruptedly. For example, for the past year, entity A had a high trust towards entity B. However, recently, entity A found that entity B lied to entity A. Consequently, there is no trust between these two entities anymore.

\item \textbf{Is trust subjective?} Although Fog networks are formed with a wide range of objects or devices, its security requirements vary from object to object or device to device. So, their trust properties are different, which is carried out more importantly over other properties. Having different types of trust policies for different objects, the trust will be subjective.

\item \textbf{Is trust transitive within a context?} Following subjective issues, each device has a distinct security policy of its own. That is, if device A trusts device C, then device A may trust any device that device C trusts in the same context. However, this concludes that the trust might be explicit and difficult to be measured.

\item \textbf{Is trust asymmetric?} Trust is an asymmetric relationship in nature. Being asymmetric in nature, trust is contrary to non-mutual relationships. It means that if device A trusts device B, we must not suggest that device B trusts device A.

\item \textbf{Is trust context-dependent?} Context is significant in terms of Fog computing~\cite{kraemer2017Fog} and at the same time, it is significant in terms of trust computing as well. Suppose, we might trust a friend to keep a secret, but not to keep our money with him. The same scenario can be applied in the Fog environment. One Fog device can be trusted to accomplish a particular task for a client in the Fog environment, but for another task, it may not trust the same Fog device. Therefore, in this situation trust needs to be context-dependent.
\end{itemize}

\subsubsection{Trust Evaluation Models}
Although Fog computing is vulnerable to any sort of illegitimate entity, it is important to ensure an effective and secure trust model that is compatible with trust computation in Fog computing.

While trust is classified amongst the imperative security requirements in Fog, there is quite a limited range of studies in the field. Most of the studies have just concentrated on the field of cloud computing.

Till now, there is no strongly recommended trust model for Fog computing, but we can enumerate already existing trust models from IoT and cloud computing. In this section, we are going to discuss a few renowned trust models which are competent for Fog computing.


\begin{itemize}

\item \textbf{Reputation-based:} The reputation-based trust model~\cite{josang2007survey} is broadly applied in peer-to-peer (P2P), e-commerce services, social media, and user reviews. Occasionally, the fame of a service provider is beneficial to select amongst diverse service providers. Damiani et al.~\cite{damiani2002reputation} demonstrated a reputation system model for P2P networks by applying a distributed polling algorithm to evaluate the consistency of the model. As this model sturdily relies on a general view, it is not appropriate in Fog computing as the nature of the end devices is dynamic. Moreover, Abhijit et al.~\cite{article10924} introduced a trust-based model to provide application layer security that can deal with the issues of user privacy, integrity and authentication. Hence, it will function as a trust-related safeguard in the Fog ecosystem for IoT related applications.



\item \textbf{Plausibility-based:} Soleymani et al.~\cite{soleymani2017secure} proposed an experienced and plausibility-based fuzzy trust model to secure a vehicular network. In a vehicular network application, it is significant to establish a trust to keep integrity and reliability. Hence, in vehicular environments, a secure trust model can handle the uncertainty and risks originating from defective information. Eventually, there are also several trusted models~\cite{yi2015security} regarding special hardware.

\item \textbf{Trusted execution environment (TEE):} TEE is an isolated environment, which guarantees the confidentiality and integrity of code and data by executing in the secure area inside a processor.

\item \textbf{Secure element (SE):} SE stores sensitive information securely and run the apps in a microprocessor chip to protect the data and application from malware attacks. 

\item \textbf{Trusted platform module (TPM):} TPM stores the host identification key pairs, which are used for hardware authentication inside a specialized chip. The data inside this chip cannot be accessed by software.

\end{itemize}


\subsubsection{Attacks on Trust Computation Environment}
In Fog computing, while Fog nodes and clients are communicating with each other, they must establish a connection with greater trust value in the Fog network. For Fog nodes and clients, the highly trusted nodes and clients will be selected and accepted frequently rather than Fog nodes and clients with lower trust. It helps to speed up the overall performance of the Fog network~\cite{guo2017survey}. Malicious intruders will impersonate their nodes as highly trusted nodes, so that, they can gain the possibility of compromising a network. In this segment we are going to define several types of attacks which might occur in the Fog network:

\begin{itemize}
\item \textbf{Self-promotion attack (SPA):} in SPA attack, the malicious Fog nodes increase their trust values to impersonate themselves as the highest trusted nodes.
\end{itemize}

\begin{itemize}
\item \textbf{Bad-mouthing attack (BMA):} this attack works by spreading fictitious information. Several malicious Fog nodes work together to provide depraved suggestions about a decent Fog node, which will damage the fame of those nodes. This is a form of a collision attack, and it happens when numerous malicious nodes come together to spread false information.
\end{itemize}

\begin{itemize}
\item \textbf{Ballot-stuffing attack (BSA):} this attack is similar to the collusion attack, where a malicious node transfers decent suggestion regarding another wicked node to raise the fame of the malicious nodes.
\end{itemize}

\begin{itemize}
\item \textbf{Opportunistic service attacks (OSA):} after assuming that the fame has been lowered down by the Fog node, it can achieve a great service to retrieve its reputation.
\end{itemize}

\begin{itemize}
\item \textbf{On-off attack (OOA):} A malicious Fog node can provide bad and good services simultaneously to avoid being rated as a low trusted node. The OOA attacker can also behave differently with different neighbors to achieve an inconsistent trust opinion of the same node.
\end{itemize}

In accordance with the study above and based on different issues, we have illustrated, a summary table on the existing related research works related to trust issues are shown in Table~\ref{sumtrust}.

\begin{table*}[htbp]
    \centering
    \small
    \caption{The summary of Trust Issues in Fog environment from major survey papers}
    \label{sumtrust}
    \begin{tabular}{L{2.5cm}R{7cm}C{6.5cm}} 
        \toprule
        \textbf{Reference Paper} & \textbf{Highlights/Objectives} & \textbf{Achievement and Limitation}\\ \midrule
        
Rauf et al.~\cite{rauf2018security} & 
\begin{itemize}
\item Propose a risk-based trust model for the IoT environment.
\item Dynamic domain adaptive security solution.
\item Parameters such as availability, reliability, response time, etc. used.
\item Direct and indirect observation also used for trust computation.
\end{itemize} &

\begin{itemize}
\item The system can compute trust as well as compute risk levels of the system.
\item Layer-wise Various attacks discussed.
\item The system will provide trustworthy information forwarding decision on the basis of trust and risk values.
\end{itemize}

\\ \midrule
Wang et al.~\cite{wang2018novel} & 
\begin{itemize}
\item Performed a Fog-based hierarchical trust mechanism.
\item Solve resource consumption problems.
\item Able to monitor the trust state of the whole network.
\item Detect and recover data attacks and misjudgment nodes respectively.
\end{itemize}&

\begin{itemize}
\item Reduce consumption of the energy by the network.
\item Ensure the state of trust for network and edge nodes.
\item Detect some attacks of hidden data.
\item Recover misjudgment nodes.
\end{itemize}

\\ \midrule
Rahman et al.~\cite{rahman2018find} & 
\begin{itemize}
\item A broker based trust mechanism approach in Fog.
\item Deliberate the trustworthy Fog service.
\item Request matching algorithm has been used.
\end{itemize}&

\begin{itemize}
\item Applies fuzzy logic for trust evaluation.
\item Able to performed dynamic trust operation.
\item Simultaneously maintained a trust relationship.
\end{itemize}

\\ \midrule
Soleymani et al.~\cite{soleymani2017secure} & 
\begin{itemize}
\item Secure trust establishment among vehicles.
\item Fuzzy trust scheme based on plausibility and experience.
\item Demonstrated a series of security checks.
\end{itemize}&

\begin{itemize}
\item Can deal with uncertainties and risks.
\item Detects faulty nodes and malicious attackers.
\end{itemize}

\\ \midrule
Yuan et al.~\cite{yuan2018reliable} & 
\begin{itemize}
\item Reliable and lightweight trust evaluation mechanism.
\item More feasible against bad-mouthing attacks.
\item Employ fusion of Multi-source feedback information.
\item Used objective information entropy theory.
\end{itemize}&

\begin{itemize}
\item Suit for IoT edge computing on a large scale.
\item Facilitates low-overhead trust computing algorithms.
\item Trust factors are weighted manually or subjectively.
\item Gained computational efficiency and reliability.
\end{itemize}

\\ \midrule
Dang et al.~\cite{dang7946404} & 
\begin{itemize}
\item A data protection scheme has been for Fog computing.
\item Dynamic and can handle mobility management service.
\item Introducing Fog-based region verification and privacy-aware role-based access control techniques.
\end{itemize}&

\begin{itemize}
\item Able to deliberate up-to-date location services.
\item Efficient and feasible scheme.
\end{itemize}

\\
        \bottomrule
    \end{tabular}
\end{table*}
\subsection{Privacy in Fog Computing}
Privacy is a key issue in any distributed environment. Across available literature, there are many mechanisms, which have been proposed to ensure the privacy of the data, such as encryption and hashing. However, these techniques are not suitable in the Fog, because it affects the latency and time to process the application. The remaining part of the section discusses in detail the privacy assurance issues.

\textbf{Privacy Assurance:} Privacy assurance helps to preserve any private information, such as data, user, usage, locations, devices, network from unauthorized access~\cite{aghasian2018user}~\cite{fu2016privacy}. In Fog Computing, all the data used comes from various sources like IoT devices, wireless networks as well as cloud networks. These data might be meaningful or meaningless, but we need to preserve it. Thus, appropriate privacy assurance can be treated as a substantial security issue in the Fog environment. There are also a few encounters ascends for privacy preservation, as the nodes are located adjacent to the end-users and they can gather sensitive information~\cite{yi2015security}.

\subsubsection{\textbf{Privacy dimensions}} Fog computing is used to work with sensitive information which is generated from several sources. For securing these types of sensitive information, privacy is one of the most significant problems in Fog computing. There are lots of privacy issues that arise in the Fog environment. In the following section, we are going to describe Fog computing privacy issues from a different perspective:

\begin{itemize}
\item \textbf{Users Privacy:} usually Fog computing consists of a large collection of IoT enabled devices which are connected through sensor or wireless network. Therefore, IoT devices are used to generate sensitive data at the user level and upload it to Fog nodes for further processing. Sensitive data such as personal data, home-automated data, business data, health data, etc. By analyzing all this sensitive information, an intruder can reveal a lot about a user's personal data and gain adequate knowledge.
\end{itemize}

\begin{itemize}
\item \textbf{Data Privacy:} as we already know that, Fog node works at the edge plane of the network and it generally collects sensitive data that is generated by various sensing and end-user devices. Hence, Fog nodes are managed by third parties. So, when all the unprocessed data are being aggregated in the Fog layer, there might be a chance to (compromise, alter, miss-match, etc) the data. Under such circumstances, we need to indemnify the privacy of these data. Usually, Fog nodes send requests to the end-users to send their private data to them, in order to further process it, store it temporarily, and finally, send data to the cloud for permanent storage. Therefore, users will not have control over the data where all the access and control will be transferred to the Fog or cloud service providers. Under such circumstances, service providers or malicious insiders can manipulate the stored data. This signifies a privacy issue to the user's data.
\end{itemize}

\begin{itemize}
\item \textbf{Usage Privacy:} this privacy issue arises when a Fog client can avail of the required Fog services. For example, in a smart grid system, the reading of the smart meter reveals masses of information of a smart-house such as at the TV on and off time or when the home is vacant, which certainly brings privacy breaches for users~\cite{aghasian2017scoring}.
\end{itemize}

\begin{itemize}
\item \textbf{Network Privacy:} wireless connectivity is comprehensive under the control of IoT as well as other edge devices in a Fog computing environment. It is a big matter of concern, as wireless connectivity is prone to network privacy attacks. The maintenance cost is correlated with the Fog nodes as it is positioned at the edge of the Internet, where network configurations are established manually~\cite{yi2015security}. The breaches private data which is an important issue while using Fog networks. The end-users share resources which contribute to Fog processing. Due to this, information that is more sensitive is collected by the Fog network as compared to a remote cloud. To overcome these issues, an encryption scheme like HAN (Home-Area Network) might be useful.
\end{itemize}

\begin{itemize}
\item \textbf{Location Privacy:} in the Fog environment, the location privacy denotes to the protective techniques for breaches related to the client's location. While the client uploads its responsibilities to the closest node, the uploaded node can assume that the client is contiguous and far away from other Fog processing devices. Therefore, if a client in the Fog environment uses multiple Fog application services from multiple locations, it may reveal its track directly to the Fog nodes, to avoid collision amongst the Fog nodes. As Fog nodes are vulnerable to potential attacks, It is easy to compromise the privacy by having the location credentials of the Fog clients. If the Fog clients are attached to an object or a person, then the location privacy is at risk. Whenever a Fog client frequently selects its closest Fog node, the node can certainly identify if the client is using the resources residing nearby.
\end{itemize}

\subsubsection{Privacy Preservation}
In Fog computing, it is used to collect and process user personal data which is desirable. So, it is evident that a proper privacy-preserving and security mechanism is required to cope up with the Fog computing environment. As we know, Fog computing consists of various devices that are connected to IoT as well as Cloud. So, we should apply privacy-preserving techniques between cloud and Fog to maintain data privacy because both Fog and cloud devices are resourceful and have adequate storage and power. On the contrary, IoT devices have limited resources. So, it's a difficult task to implement privacy-preserving techniques between the Fog and IoT devices. It is significant because the users and users' may be concerned about their data which is sensitive~\cite{koo2016hybrid}. Different privacy preservation techniques, methods and schemes are proposed across many scenarios, including cloud~\cite{cao2014privacy}, wireless network~\cite{qin2014preserving}, smart grid~\cite{rial2011privacy}, health-care systems~\cite{al2017security}, and online social network~\cite{novak2014near}.

\begin{itemize}
    
\item \textbf{Homomorphic encryption:} There is a method for privacy- preservation, which is homomorphic encryption (it is a method for operating encrypted data without decrypting it), that can be implemented to retain the privacy of transmitted data without decryption across local gateways~\cite{lu2012eppa}.

\item \textbf{Differential privacy}~\cite{dwork2011differential}: is to assure the privacy of random individual entries in the statistical data set. Although its computational overhead for such function is a big issue in Fog computing, it needs to be assiduous about the efficiency of the method.  

\item \textbf{Identity obfuscation:} There is a renowned technique called identity obfuscation technique~\cite{wei2012mobishare}, where the Fog node is able to recognize the Fog client is close by, but it cannot recognize the Fog client. As such, identity obfuscation is a technique for preserving location privacy, as it has many methods inwardly. There is an elementary method to preserve the location privacy of the Fog client, whereby this client is allowed to upload the data between diversified Fog nodes. This method is not efficient, because it would waste Fog resources and enhance the latency. As we already know, the Fog client can choose its nearby Fog node to upload its data, so the Fog node is able to identify that the Fog client is residing nearby, which helps to get the Fog client's location credentials.

\item \textbf{Trusted third party:} Wei et al. ~\cite{wei2012mobishare} demonstrated a method, where a trusted third party (TTP) generated a fraudulent ID for each Fog client. As a matter of fact, it is not necessary that the Fog client has to choose a node which is nearby, in spite of that it can choose any nodes on the basis of a stipulated set of criteria such that the reputation, latency or load balancing is not affected. In this scenario, the Fog node can recognize the Fog client's rough location but cannot detect it exactly. In addition, there could be a scenario whereby a Fog client uses resources from multiple Fog nodes or the location of the client can be squeezed into a small region. As such, the location of the client must be within the coverage of several Fog nodes. According to the described scenario, the authors~\cite{gao2013location}, used a method to preserve location privacy.

\item \textbf{Data partitioning:} Another probable method could be effective for preserving user privacy by partitioning the data into multiple Fog nodes. The usage pattern is another privacy concern when clients are using Fog services. In this scenario, privacy-preservation techniques have been suggested in smart metering~\cite{mclaughlin2011protecting,rial2011privacy}, but we cannot apply these mechanisms in Fog computing directly, because there is no TTP (i.e., smart meters in the smart grid) or no backup device. The Fogging device can accumulate the list of tasks for user usage. The creation of bogus tasks by the clients and uploading them to multiple nodes is one possible solution while hiding actual tasks from the bogus ones. However, this solution may not be operational as it raises the client's expense and wastes resources. 
\end{itemize}

According to the discussion above and based on different criteria for privacy-preservation, it has summarized into the Table \ref{sumpri}.

\begin{table*}[htbp]
    \centering
    \small
    \caption{The summary of privacy issues in Fog environment from major survey papers}
    \label{sumpri}
    \begin{tabular}{L{2.5cm}R{2.5cm}C{6cm}C{5cm}} 
        \toprule
        \textbf{Reference Paper} & \textbf{Privacy Issues} & \textbf{Highlights/Objectives} & \textbf{Performances and Achievements}\\ \midrule
Wang et al.~\cite{wang2018anonymous} & Data Privacy, Identity Privacy & 
\begin{itemize}
\item Fog based public cloud computing.
\item The idea of anonymity and secure aggregation techniques used.
\item Provide identity and data privacy.
\item Performed pseudonyms and homomorphic encryption techniques.
\end{itemize} &
\begin{itemize}
 
\item Performed computation and communication effectively and efficiently.
\item Can save the communication bandwidth.
\end{itemize}

 \\ \midrule
Yang et al.~\cite{yang2018position} & Location privacy, Location verification & 
\begin{itemize}
\item Introduced secure positioning protocols by preserving the location privacy.
\item Position based advanced cryptographic protocols have been introduced, which preserve the location privacy.
\end{itemize}&
\begin{itemize}
\item Privacy is gained without utilizing additional computational overhead.
\item The system is as efficient and quite practical in practice.
\end{itemize}

\\ \midrule
Kumar et al.~\cite{kumar2016Fog} & Location Privacy, Data Privacy    & 
\begin{itemize}
\item Data confidentiality and location privacy are focused on.
\item Discussed how to access user data.
\item The misconceptions about the rights of users were discussed.
\item The concept of a decoy method with some incorporation for data and location privacy.
\end{itemize}&
\begin{itemize}
\item The concept of decoy method for data and location privacy has been discussed.
\item Different attackers and their interest in a user's private data was also discussed.
\end{itemize}

\\ \midrule
Liu et al.~\cite{liu2018secure} & Location privacy, Identity privacy & 
\begin{itemize}
\item Fog based vehicular ad-hoc network (VANET)
\item Secure and intelligent traffic light control system using Fog.
\item Location Based Encryption (LBE) and Cryptographic computational Diffie–Hellman puzzle has been used.
\end{itemize}&
\begin{itemize}
\item Reduce the computation and communication overhead.
\item Traffic light may efficiently verify the authenticity of the vehicles.
\item Fog device friendly and is able to defend the Denial-of-Service (DoS) attack.
\end{itemize}

\\ \midrule
Lu et al.~\cite{lu2017lightweight} & Device Privacy, Data Privacy &        
\begin{itemize}
\item Employing lightweight privacy-preserving data aggregation method, for Fog and IoT systems.
\item The homomorphic Paillier encryption, Chinese Remainder Theorem, and one-way hash chain techniques have been applied.
\end{itemize}&
\begin{itemize}
\item Performed efficiently and aggregated hybrid IoT devices data into one.
\item Supported fault-tolerance(FT).
\item Prevents false data injection attack by filtering injected false data at the network edge level.
\item Computation and communication costs are very low. 
\end{itemize}

\\ \midrule
Qin et al.~\cite{qin2014preserving} & User's privacy, Network Privacy, Data Privacy & 
        
\begin{itemize}
\item Preservation of the privacy of the end user's over a radio network.
\item Techniques used include commitment schemes along with zero-knowledge proof and random-checking monitoring to preserve the privacy of the end user and to protect the data flow over the radio network.
\end{itemize}&
\begin{itemize}
\item Provides user's privacy, data security and network privacy in the Fog computing environment
\item Efficiency and accuracy is unpredictable in the Fog computing environment.
\end{itemize}

\\
        \bottomrule
    \end{tabular}
\end{table*}


\subsection{Authentication in Fog Computing}
\begin{figure}[!t]
    \centering
    \includegraphics[width=3in]{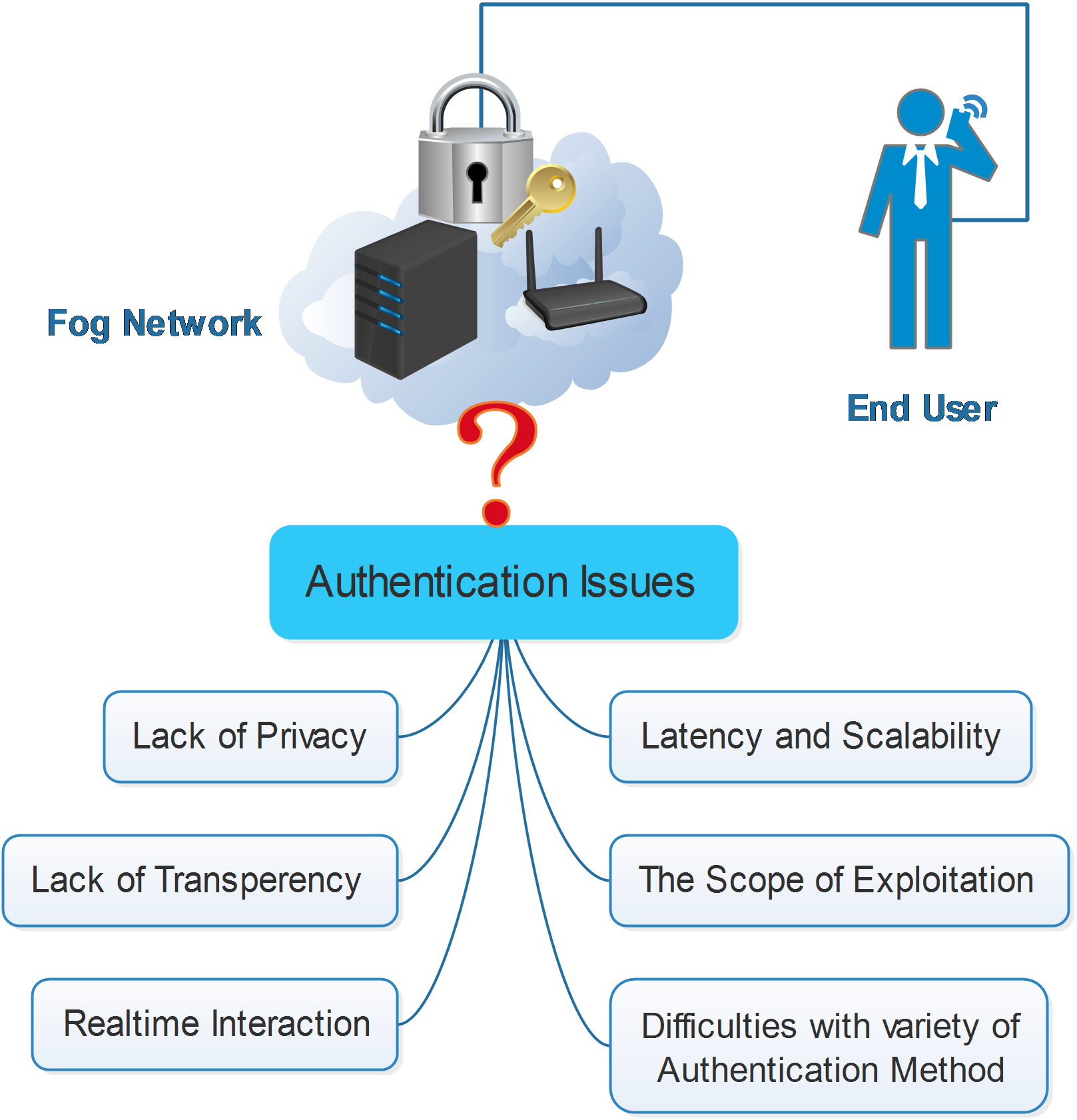}
    \caption{Authentication issues arises in Fog Computing.}
    \label{fig_auth}
\end{figure}

Authentication helps to verify a user's identity by verifying if a user's credentials match with the information in a database via the authentication server. In the context of Fog computing, authentication ensures and confirms an end user's identity. This helps ensure that only legitimate end users can have access to the Fog nodes who have met all the requirements to be authenticated as an end-user. Authentication is one of the five pillars of Information Assurance  (IA)~\cite{ahmadizadeh2015automated}. In Fog computing, authentication of the end user's devices permitted to Fog services is a significant requirement in the Fog network. In order to obtain the Fog services from the Fog infrastructure, an end user's device must be authenticated to be a part of the Fog processing infrastructure by authenticating itself.  Whereas it is also essential to defend against the access of unauthorized entities. Fig. \ref{fig_auth} shows the authentication issues in Fog computing.


With the higher number of internet-enabled devices, authentication is getting more and more vital to permit secure communication for IoT applications and home automation. Almost any object (entity) may be addressable and be capable to exchange information over the network. Thus, it is significant to comprehend that each device or application can be potentially an intrusion point in the environment. So, it is mandatory to ensure a strong authentication mechanism for each device or application in the Fog network system.

Although Fog computing eliminates many difficulties compared to primitive cloud computing, it also provides excellent services such as mobility, geo-distribution, heterogeneity, real-time processing, etc. Similar to Cloud computing, Fog computing also faces new security challenges. Due to heterogeneity and interaction of third party authorities in the Fog computing system, it leads to an increase in the scope of security breaches. In such a case, there might occur various renowned attacks (e.g. data loss, account traffic hijacking, man-in-the-middle attack, denial of service attack, malicious insider attack, etc). Therefore, it is a significant issue to think about secure Fog networks by ensuring the security mechanism in every stage. In that case, authentication plays a key role in protecting the Fog network. Therefore, ensuring proper authentication mechanisms would be a suitable solution to prevent such attacks. As Fog computing is used to provides various services with low latency and cooperate with the edge devices as well as cloud systems, by providing any authentication mechanism, there might be a chance to raise critical issues such as latency, scalability and efficiency which needs to be handled according to the demands of the Fog computing environment.

\subsubsection{Authentication Factors in Fog computing}
The authentication factor refers to attributes or data that can be considered to authenticate user access to a system. A legacy security system has a few authentication factors such as the knowledge factor, which is something users know, the possession factor which is something a user has and the inherent factor which is something the user is.  In recent years, other authentication factors have been added - location factor and time factor, along with the old authentication factor which are as follows:

\begin{itemize}
\item \textbf{Knowledge Factor:}
the knowledge factor is any credentials that consist of information that the user holds, such as Username, Password, Personal Identification Number (PIN) and answers to the secret questions~\cite{aghasian2018privacy}.
\end{itemize}

\begin{itemize}
\item \textbf{Authority Factor:}
the authority factor would be any credentials that the user can own and carry with them, such as hardware devices like a mobile phone or a security token.
\end{itemize}

\begin{itemize}
\item \textbf{Inherent Factor:}
the inherent factor is generally based on biometric identification (fingerprints, facial, retina).
\end{itemize}

\begin{itemize}
\item \textbf{Location Factor:}
the location factor itself cannot usually refer to authentication, but it can be used with other factors. For example, a legitimate user normally can access a system from home or office in any organization's home country. An attacker will try to access that system from a remote geographical location. With the help of a location factor, the system can prevent illegitimate user authentication into a system or network.
\end{itemize}

\begin{itemize}
\item \textbf{Time factor:}
similar to the location factor, the time factor can be used as a supplement with other factors. It can be used together with the location factor. For example, an authorized user can have access to a system in a specific time period in an organization's home country. On the other hand, an illegitimate user tries to access that system from a remote geographical location of another country. Therefore, the authentication would be rejected based on the time and location factor.
\end{itemize}

\subsubsection{Authentication Measures in Fog Computing}

\begin{itemize}
\item \textbf{Lack of Transparency:}
The existence of SLA between a Fog or cloud service and an end users is a vital issue in order to establish trust. Although many SLAs have clearly defined the privacy over the user's sensitive data, users are unable to trust them in how the data is being governed. Hence, the SLA verification gets limited when the service is being directly used in the Fog layer by the end users and a small organization, which should be monitored by a licensed third-party through SLA verification. There might be a lack of transparency that permits the users to monitor their own data in the Fog or cloud system.
\end{itemize}

\begin{itemize}
\item \textbf{Real-time Interaction:} Fog nodes and end users interact with a huge number of devices simultaneously. Different services needs different authentication mechanisms where if the process takes a huge time to authenticate, it would be a challenging task with respect to real-time interaction.
\end{itemize}

\begin{itemize}
\item \textbf{Latency and Scalability:} In accordance with the rapid growth of user devices and services, it is an ambitious task to guarantee the efficiency of the authentication mechanism. Whenever the latency of the authentication process is high and incompatible with the service, scalability is a big concern.
\end{itemize}

\begin{itemize}
\item \textbf{The scope of Exploitation:} In the context of Fog or cloud system, there is a diversified authentication mechanism for various services. These authentication methods can be compromised or exploited by the attacker and the attacker can appear to have gained administrative level access due to the deficiency in the authentication mechanism. There might be a chance to breach the security of data, devices as well as the Fog network system.
\end{itemize}


\subsubsection{Authentication Techniques in the Fog Environment}
Generally, users need to use various services simultaneously. Therefore, they need to use different authentication methods for different services where the performance of the authentication methods are different in the context of latency, efficiency and scalability. On the other hand, the user faces lots of difficulties to maintain access credentials for multiple services. Authentication is the most significant issue for the security and privacy of Fog computing. An authentication mechanism that is not secure might cause harm for the cloud, Fog and end user's devices, which is one of the main security concerns for Fog computing ~\cite{stojmenovic2016overview} as well. Therefore, different authentication techniques have been proposed for elevating security mechanisms in the Fog or cloud computing, but each authentication method has come up with its own dominance and limitations. In this subsection, a few traditional authentication techniques and their limitations as well as drawbacks according to the Fog environment has been described. We also described a few proposed solutions which meets with the Fog computing criteria.

\begin{itemize}
\item \textbf{Password Based Authentication:} In password authentication, the user must first give a password for every service, and the system administrator must keep track of all usernames and passwords on the server. Password Authentication is performed by accepting a key and password for allowing a user into local and remote systems. Password authentication can be categorized depending on its strength as weak authentication, stronger authentication, and inconvenient authentication~\cite{mohammedpassword}. Therefore, password-based authentication has several applications and it is deployed in cloud computing~\cite{tsai2009efficient}~\cite{lu2008simple}~\cite{kumar2010enhanced}, but it will face numerous drawbacks and limitations when it is considered for Fog computing:
\begin{itemize}
\item It takes an extensive computation to process. it's challenging due to the limited end device resources.
\item In the Fog network, end users frequently communicate with various Fog nodes from different Fog environment. Therefore, it is inappropriate to keep a password for each Fog node. In addition, it is not a good concept to set the most used password for each Fog node.
\item –	Usually, a password does not provide high security because of numerous attacks~\cite{lee2013guessing}, for example, vulnerability to off-line dictionary attacks.
\end{itemize}
\end{itemize}

\begin{itemize}
\item \textbf{PKI Based Authentication:} public key infrastructure (PKI) based authentication creates and upholds a reliable networking environment by offering certificate and key management services that permit encryption and digital signature abilities between applications all in a way that is transparent and easy to use. PKI offers confidentiality, authentication, integrity (CIA) and non-repudiation of the exchanged messages.  In~\cite{stojmenovic2014Fog}, the authors described security issues and focused on authentication issues at various levels of the Fog computing environment. Therefore, the traditional PKI-based authentication scheme is not effective in the context of Fog computing due to the poor scalability. In addition, the allocation of public keys can be weighty due to the enormous scale of Fog nodes and end users. Another drawback is that, if the private keys cannot be well preserved, the security will be ruined.

\end{itemize}

On the other hand, the Diffie-Hellman~\cite{fadlullah2011toward} key exchange based authentication scheme is not compatible with the Fog environment due to its slow and extensive computations.

Balfanz et al.~\cite{balfanz2002talking} demonstrated a user-friendly, cheap and secure method to resolve the authentication issue for a wireless networks based on pre-authentication of location-limited channel. Likewise, Nearfield communication (NFC) is used in Cloudlet to simplify the authentication process~\cite{bouzefrane2014cloudlets}. Ibrahim et el.~\cite{ibrahim2016octopus} proposed a secure mutual authentication method for the Fog environment, that allows authenticating any Fog user with the Fog nodes mutually in the Fog network. The authors~\cite{manzoorsecure} proposed a method based on the multi-Tier authentication scheme to Secure Login in Fog Computing. The authors~\cite{vishwanath2016security} mentioned that Advance Encryption Standard (AES) is a compatible encryption algorithm for the Fog computing environment as it needs low hardware resources and fewer computations. The authors~\cite{stojmenovic2016overview} demonstrated that the end user devices can initiate spoofing attacks and are prone to data tampering which can be preserved with the aid of PKI, DiffeHellman key exchange and monitoring by Intrusion detection techniques. Finally, the authors adviced that the chances of such attacks can be prevented by deploying a secure authentication mechanism between the Fog platform and the end users.

\begin{itemize}
\item \textbf{Biometric Authentication:}
is a technique of user identity verification based on various biological inputs through scanning or analysis of some parts of the body. Biometric scanners scanning a user's physical biometric characteristics such as fingerprint, voice recognition, iris scan, face recognition, etc. Generally, biometric authentication takes place to manage access to digital or physical resources. Biometric authentication is an upcoming technology and is already rapidly deployed in mobile computing as well as cloud computing using fingerprint authentication, face authentication, keystroke-based authentication or touch-based authentication~\cite{yi2015security}. On the other hand, biometric authentication techniques comparatively take a huge execution time and its security level remains constrained when high-level security is required~\cite{ibrahim2016octopus}. Therefore, in accordance with the Fog computing environment, applying biometric-based authentication techniques would be a suitable solution. Although still, it has a lot of limitations and drawbacks - it takes more computational time during the process of execution and it provides constrained levels of security when high-level security is required. Therefore, to consider biometric based authentication for Fog computing is still a research issue~\cite{yi2015security}.

\end{itemize}

In accordance with the study above, and based on different issues of authentication, this has been summarized in Table \ref{sumAuth}.

\begin{table*}[htbp]
    \centering
    \small
    \caption{The summary of Authentication Issue in Fog environment from major survey papers}
    \label{sumAuth}
    \begin{tabular}{L{2.5cm}C{7cm}C{6.5cm}} 
        \toprule
        \textbf{Reference Paper} & \textbf{Highlights/Objectives} & \textbf{Performances and Achievements}\\ \midrule
        
Ibrahim et al.~\cite{ibrahim2016octopus} & 
\begin{itemize}
\item An efficient and secure mutual authentication method for the cloud-Fog-edge system architecture.
\item Required to store one master secret key.
\item Does not need extra overheads such as re-initialization or re-registration process.
\end{itemize} &
\begin{itemize}
\item Required to perform fewer hash invocations and symmetric encryptions/decryptions.
\item In addition, simple countermeasures have been introduced.
\item Suitable and can be deployed efficiently to the Fog user's smart device/card.
\end{itemize}

\\ \midrule

Wazid et al.~\cite{wazid2019design} & 
\begin{itemize}
\item Fog devices security can be ensured through key management and authentication schemes.
\item Performed efficient and lightweight operations.
\item Bitwise exclusive-OR (XOR) and One-way cryptographic hash function techniques have been considered.
\item Demonstrated using formal security verification. 
\end{itemize}&
\begin{itemize}
\item Performed low computation and communication overheads.
\item Ensure high security compare to another existing method.
\end{itemize}

\\ \midrule

Dsouza et al.~\cite{dsouza2014policy} & 
\begin{itemize}
\item Introduce a policy-based resources management in Fog network.
\item Support interoperability and secure collaboration among various resources in Fog system.
\end{itemize}&
\begin{itemize}
\item Server authentication, device authentication, data migration authentication and instance authentication has been observed for the secured Fog computing environment.
\end{itemize}

\\ \midrule

Alharbi et al.~\cite{alharbi2017secure} & 
\begin{itemize}
\item Ensure secure communications among the various IoT devices.
\item Performed challenge-response authentication technique.
\end{itemize}&
\begin{itemize}
\item Performed effectively and efficiently.
\item It can achieve very low response latency.
\item Protects the IoT system from DDoS attacks.
\end{itemize}

\\ \midrule

Amor et al.~\cite{amor2017privacy} & 
\begin{itemize}
\item Introduces anonymous mutual-authentication amongst the Fog users and Fog servers.
\item Cryptographic and mathematical have been performed to establish the session key.
\end{itemize}&
\begin{itemize}
\item Can accomplish effectively and efficiently and improved the security and privacy in Fog network.
\item Can defend against various attacks such as man-in-the-middle attack, eavesdropping and reply attacks. 
\end{itemize}

\\ \midrule

Hu et al.~\cite{hu2017security} & 
\begin{itemize}
\item Highlighted privacy-preservation and security methods for Fog based image processing applications.
\item Data encryption, the authentication and session key agreement,  and data integrity checking such methods have been proposed.
\end{itemize}&
\begin{itemize}
\item Can perform effectively and solve the issues of integrity, availability, and confidentiality.
\item Increases a little computation and communication overhead.
\end{itemize}

\\ \midrule

Ha et al.~\cite{ha2016efficient} & 
\begin{itemize}
\item An efficient and elliptic cryptographic based mutual-authentication technique for an IoT based resource constrained devices.
\item Uses Implicit certificate and key management for secure communication and mutual authentication.
\end{itemize}&
\begin{itemize}
\item Achieved less execution time.
\item Suitable for resource constrained devices.
\end{itemize}

\\ \midrule

Gope et al.~\cite{gope2018lightweight} & 
\begin{itemize}
\item Deliberated two-factor lightweight and privacy-preserving authentication method for resource constrained IoT devices.
\item Provide resilient way of authentication.
\end{itemize}&
\begin{itemize}
\item Very efficient computational capacity.
\item Can performed robustly against malicious attacks.
\end{itemize}

\\ \bottomrule
    \end{tabular}
\end{table*}

\subsection{Access Control in Fog}
\begin{figure}[!t]
    \centering
    \includegraphics[width=3in]{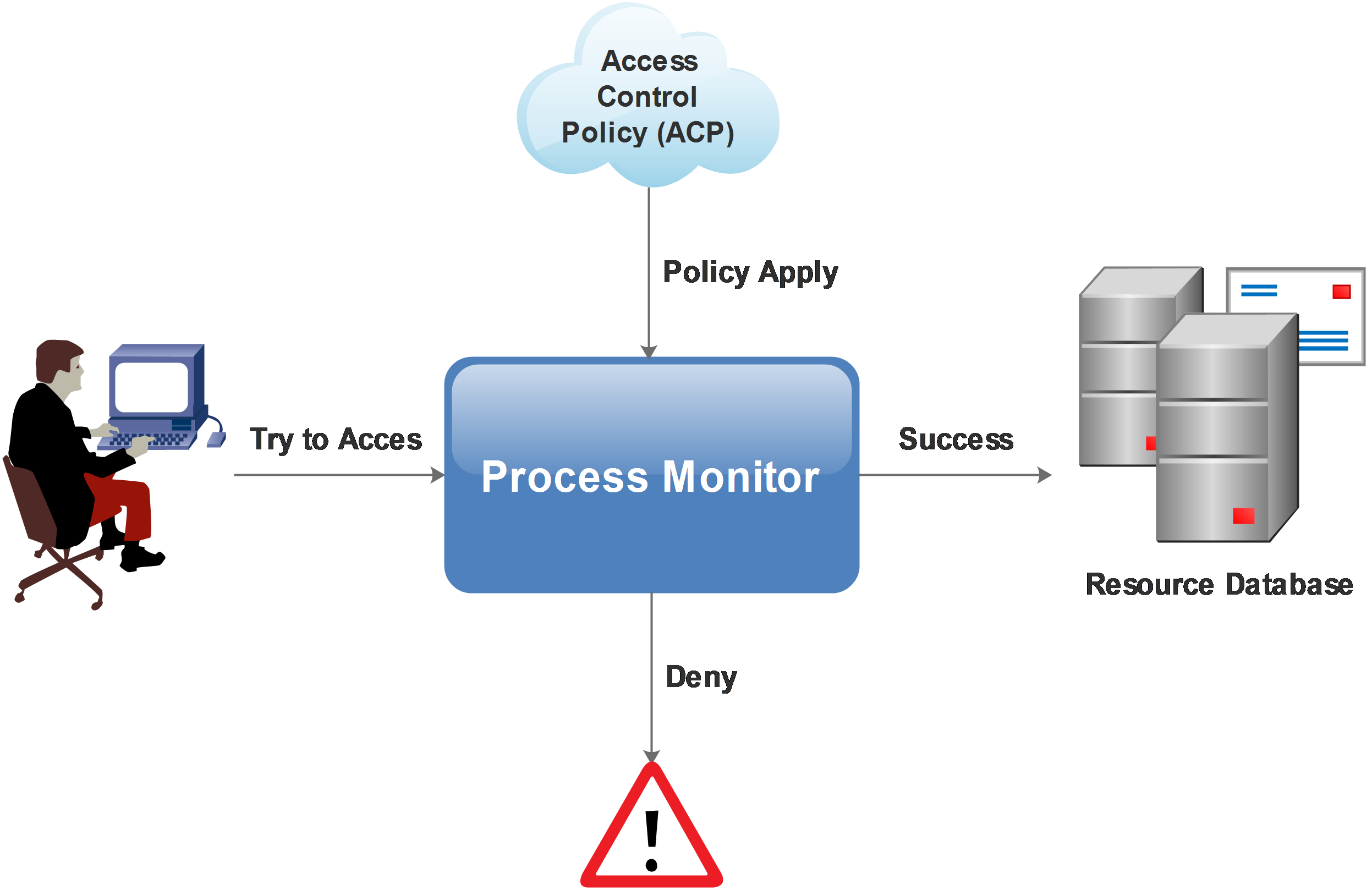}
    \caption{A Process of Access Control (AC).}
    \label{fig_Acl}
\end{figure}

Access control is a method of restrictive access to a system or to a physical or virtual resource. In computing, it is defined as a process by which users are granted privileges for retrieving information from the system, information or resources. In access control systems, individuals must have legitimate credentials before access can be granted to them. The process of access control is shown in Fig. \ref{fig_Acl}

By deploying Access Control in the Fog network system, it would be possible to conserve a user's privacy and assure both the user and system security maintain trust between the Fog, cloud service providers and users. The authors in~\cite{8291130} highlighted a few Access Control (AC) problems in the area of Fog computing and classified these problems into the following types:

\begin{itemize}
\item The users should be authenticated by the Fog or cloud system if they wanted to use the services such as storage or computation, where several  strategies must be used to control access for both services and data as well.
\item Security management is difficult to control, given the number of requirements.
\item The cloud and Fog system needs mutual access control.
\item Access control mechanism helps to prevent attacks  such as side-channel in Virtual machines (VMs).
\item Resources are very limited to both the user and Fog devices respectively.
\end{itemize}

\subsubsection{Access Control Models}
Access control is the best methods to achieve preservation within the networks, devices and systems. While it helps user's admittance in the system, access control also supports efficient data protection from various kinds of adversaries. Conventionally, access control models (ACM) are categorized~\cite{meghanathan2013review} into the following forms.
 \begin{itemize}
\item{\textbf{Discretionary access control (DAC):}}
the object's owner elects access permissions to others. These models are typically used in traditional applications of cloud and suffers from significant overhead costs in managing the multi-user environment. The second category abstract requires the need of resource-user mapping. So, compared to DAC models, this model is more flexible for distributed systems.

\item{\textbf{Mandatory access control (MAC):}}
The MAC models use multi-level security systems. Here, the administrator of the system decides who has access to the system. In a multi-level MAC model, both objects and subjects are recognized with a security level classification (i.e. top secret, secret, classified, and unclassified). The nature of Fog/cloud computing is outsourced, hence there is a need to focus on access control models which can be effectively applied in this computing environment. 

\item{\textbf{Role Based Model (RBAC):}}
Designing a model for access control is a rudimentary challenge in a large scale to secure mobile distributed applications and database systems as there is a need to provide dynamic privileges for checking systems in the environment. RBAC is a fined grain model that offers more benefits compared to previous models~\cite{vohra2018multi}, such as regulating the user's access to applications and resources by identifying the activities and the roles of users in the system~\cite{sandhu1996role}. RBAC authorizes the subject based on their responsibilities and roles of individual users within the Fog-cloud computing environment~\cite{punithasurya2012analysis}~\cite{meghanathan2013review}~\cite{8291130}~\cite{sookhak2017attribute}. Roles may vary from subject (user) to subject (user). That means in this model, the responsibility of a subject is more vital than the subject itself~\cite{langaliya2015enhancing,punithasurya2012analysis}. 

\textbf{\textit{Limitations and Drawbacks of the RBAC Model:}}

\begin{itemize}
\item The RBAC model had been developed for allocating user permissions statically.
\end{itemize}

\begin{itemize}
\item It does not consider contextual information (e.g. location, time, device constrains) and dynamic/random behavior of users.
\end{itemize}

\begin{itemize}
\item It cannot cope with dynamic segregation of duties.
\end{itemize}

\begin{itemize}
\item –	It is coarse-grained. If you have a role called administrator, then you would assign the administrator role permission to ``View employee record'' (i.e it has permissions to see all the records of employed) which denotes as an expansion of the role.
\end{itemize}

\begin{itemize}
\item It ignores meta-data of resources e.g. employee owners record.
\end{itemize}

\begin{itemize}
\item It is hard to manage and maintain within a large administrative domain. 
\end{itemize}

\begin{itemize}
\item Access reviews are painful, error-prone and lengthy.
\end{itemize}

\begin{itemize}
\item Permissions accompanying each role change or delete is based on the change of the role.
\end{itemize}

Therefore, RBAC in Fog, should ensure quicker granting access permissions and minimize the above-mentioned limitations and drawbacks.

\item {\textbf{Attribute-based Access Control (ABAC):}}
This model is one of the latest methods of managing authorization. It is a talented alternative to conventional access control techniques and has attracted consideration from both academia and the industry. Comparatively, recent developments of ABAC still leaves several unknown difficulties such as delegation, administration, auditability and scalability. 

\item{\textbf{Attribute Based Encryption (ABE):}}
This model is an encryption-based Access Control model and best suits access control problems in the Fog-cloud environment. The Attribute-Based Encryption(ABE)~\cite{sahai2005fuzzy} method categorized into two types. firstly, the encryption is based on the key policy which is known as
key policy attribute based encryption (KP-ABE)~\cite{goyal2006attribute} and secondly, the encryption is based on Cipher-text policy which is known as
Cipher-text policy Attribute-based Encryption (CP-ABE)~\cite{bethencourt2007ciphertext}.

This model can preserve data privacy and enable data owners to define a desirable set of policies directly~\cite{8291130}.
\begin{itemize}
\item {\textbf{Key Policy Attribute-based Encryption (KP-ABE):}} 
Goyal et al.~\cite{goyal2006attribute} proposed KP-ABE in the year 2006, based on the classical ABE model and uses one of many communications. This technique achieves fine-grained access control with higher elasticity to control individuals compared to the traditional scheme~\cite{vohra2018multi}.
\item {\textbf{Cipher-text Policy Attribute-based Encryption (CP-ABE):}}
CP-ABE~\cite{bethencourt2007ciphertext} was introduced as another alternative form of ABE. CP-ABE can provide fine-grained and reliable access control for cloud storage environment that is not trust worthy. Users can access data only if their attributes match the access policies associated with the data. CP-ABE works in a reverse compared to KP-ABE. In this, the  key generated is attribute user set, where the cipher text is fixed by access policy~\cite{vohra2018multi}. However, CPABE has two main drawbacks~\cite{wang2017cp}: policies are not explained using standard languages and it cannot support non-monotonic policies.
\end{itemize}

\textbf{Architecture of ABE :}
The architecture of the ABE method is categorized as centralized and decentralized as well as hierarchical~\cite{sookhak2017attribute}. 
\begin{itemize}
\item \textbf{Centralized:} In a centralized architecture, the keys will be served by a central authority center for the users.
\item \textbf{Decentralized:} In a decentralized architecture, the information will be shared by multi-authorized authorities based on the policies of various organizations. 
\item \textbf{Hierarchical:} In hierarchical architecture, the scalability and flexibility is enhanced and assists the features of one-to-many encryption for the users.
\end{itemize}

\textbf{Revocation Types of ABE:} The revocation types are categorized into two types: attribute revocation and user revocation.
\begin{itemize}
\item \textbf{Attribute Revocation (AR):} by using the AR mechanism, the attribute from the user's attributes list will be removed by the revocation controller unit.
\item \textbf{User Revocation (UR):} by using the UR mechanism, a user restricts data access via the revocation controller unit. 
\end{itemize}

\textbf{Revocation Method:} There are various revocation methods to revoke a user and attributes using the  ABE method. Proxy re-encryption, time re-keying, an update key, lazy revocation and LSSS matrix are the primary revocation methods. 

\textbf{Revocation Issue:} Deploying the ABE method in cloud storage systems to control data access brings about forward and backward revocation issues.

 \textbf{Revocation Controller:}
The revocation controller is someone who is designated to execute the user or the attribute revocation method. In general, the owner of data revokes the attributes or the user but the data owner is able to confer the revocation duties to the server or the authorized entity.


\textbf{\textit{Limitations and Drawbacks of ABE Based Model:}}
 : As we mentioned before, Fog computing extends cloud and the functionalities as well as the requirements of Fog computing, which are unique. So, the access control structure of cloud computing is not able to directly meet the requirements of Fog computing. However, researchers~\cite{stojmenovic2016overview}~\cite{li2015robust} recommended that ABE techniques suits Fog computing, but still needs to improve and meet some criteria such as fine-grained, cryptographically enforces, latency and policy management problems which needs to be re-thought and considered for further research. Although the end device or user device in Fog computing is constrained resources. Therefore, there is no need for deploying data encryption-decryption and access control mechanisms at the user level. Because the Fog devices are resourceful and used close to the end-user devices. Based on these circumstances, outsourcing access control methods would be the more appropriate solution for Fog computing. On the other hand, as we know already, Fog computing consists of a dynamic environment. Therefore, the ABE-based access control should support creating, updating, and revoking the user attributes and access structures with the management of the access policies according to the dynamic behavior of Fog computing~\cite{8291130}.


\end{itemize}

\subsubsection{Issues and Requirements for Access Control in Fog Computing}
To establish and ensure secure and efficient access control, policies must ensure confidentiality, accountability and integrity. However, due to the nature of the Fog computing environment, one should consider a few things to build a secure and strong Access Control (AC)~\cite{8291130}~\cite{salonikias2015access} which are as follows:

\begin{itemize}
\item \textbf{Computation and Communication Latency:} it indicates how long it takes for a single packet to travel from one designated node to another node. The sender considers sometimes latency as the time for sending a packet and getting an acknowledgement from the sender, where the round-trip time is taken as latency. As Fog computing is renowned for its faster accessibility, we need to ensure low-latency for providing smooth services to the end users. We can indemnify the low-latency during processing time so that the access decision can transpire within a reasonable time.
\end{itemize}

\begin{itemize}
\item \textbf{Efficiency:} efficiency is also correlated to latency. In Fog computing, there are two types of devices e.g. resource rich (Smart Power Grid, Smart City, Smart Transportation System, E-Health etc.) and resource constrained (mobile phone, smart-watch, smart-glass, etc.). The proper implementation of Access Control System in Fog computing is still a challenging issue because of it's low efficiency. If the low efficiency occurs in a continuous manner, it can result in undesirable latency, which can affects the other parts of the network.
\end{itemize}

\begin{itemize}
\item \textbf{Generality:} with the distinction of hardware and software, we need to generalize all the systems and services of Fog computing.
\end{itemize}

\begin{itemize}
\item \textbf{Data Aggregation:} in Fog computing, users are geo-spatially distributed where Fog devices are used to collect data from user devices. Therefore, it is necessary to accumulate all Fog devices closer to the end users for reducing latency. The data generated from user devices will be meaningful or meaningless but it should be handled intelligently and evenly. During the whole aggregation process, authority changes are a critical issue for data access control.
\end{itemize}

\begin{itemize}
\item \textbf{Privacy Desecration:} as it is possible to exchange data between one domain to another domain, administration of the decentralized architecture of Fog computing leads us to protect the privacy of data through Fog access control. So, it becomes a critical requirement to protect the user's data privacy.
\end{itemize}

\begin{itemize}
\item \textbf{Network Availability:} in Fog computing, network availability must be defined in such a way that when there is an issue of network unavailability, access control can also deliver the predefined level of functionality. 
\end{itemize}

\begin{itemize}
\item \textbf{Context Awareness:} when multiple operations like capturing, transferring, processing and storing are running, access control decisions should be managed competently to support all the contextual information (e.g. health condition, weather condition, temperature, time, traffic condition, etc.) [81]. 
\end{itemize}

\begin{itemize}
\item \textbf{Scalability:} scalability is to facilitate the services according to the needs of the end users. In access control, scalability will provide the services that can grow or shrink according to the end user's level of capacity. For scalability, the CloudPolice~\cite{cloudpolice} have proposed a distributed solution, in which hypervisors are responsible for the communication with each other to install access control states.
\end{itemize}

\begin{itemize}
\item \textbf{Resource Restriction/Constraints:} in Fog computing, the user or the edge resources are limited. So, it becomes tough to implement access control for Fog computing.
\end{itemize}

\begin{itemize}
\item \textbf{Policy Management:} it is an integral part of Fog computing architecture. So, the access control model needs to
be capable to support creating, invoking, releasing, and deleting policy management. Dsouza et al.~\cite{dsouza2014policy} developed a policy-driven security management framework, which is capable to support secure communication and resource sharing in the Fog environment.
\end{itemize}

\begin{itemize}
\item \textbf{Accountability:} in Fog computing, it is significant to keep track of the suspicious activities of intruders. These tracks keeping should be handled intuitively across the administrative domains.
\end{itemize}

\subsubsection{Access Control Domains}
: In the Fog computing arena, for defining access control system the contextual domains are 1. Fog to Edge, 2. Fog to Fog, 3. Fog to Cloud. While edge devices are communicating and sending data to Fog devices during the time that the Fog device uses to process all the data in such a way, so that, if the necessity arises, it can send all the processed data to the nearest Fog devices. When the issues for storing data arise permanently, Fog devices are able to send all the data to a data warehouse or cloud storage. Therefore,  process/store identity and access data in the Fog/cloud computing by first ensuring secure Fog/cloud access control. 
Ensuring access control in the cloud/Fog environment is a crucial technique to enhances the user security. In this scenario, end-user/data privacy, faster communication and computation, network and communication security, etc. Such requirements shall be applied for the above-mentioned domains to enable the proper access control system. For this, all the primordial access control models are being advanced accordingly.

In accordance with the above study, and based on different access control method, it has been summarized into Table \ref{sumAccess}.

\begin{table*}[htbp]
    \centering
    \small
    \caption{The summary of Access Control Issue in the Fog environment across major survey papers}
    \label{sumAccess}
    \begin{tabular}{L{2.5cm}C{7cm}C{6.5cm}} 
        \toprule
        \textbf{Reference Paper} & \textbf{Highlights/Objectives} & \textbf{Performances and Achievements}\\ \midrule

Zhang et al.~\cite{zhang2018efficient}&
\begin{itemize}
\item A promising CP-ABE based access control for a Fog computing environment.
\item Outsourcing and attribute update capability.
\item Encryption and decryption are outsourced.

\end{itemize}&
\begin{itemize} 
\item Perform heavy computation operations of encryption and decryption within a very small and constant time period.
\item Less computation cost and efficient attribute update.
\item Suitable for resource-constrained IoT devices.
\end{itemize}

\\ \midrule
Vohra et al.~\cite{vohra2018multi} & 
\begin{itemize}
\item Fog based decentralized Multi-Authority attribute based data access control.
\item Also based on CP-ABE method.
\item Performs fast offline-online encryption and partial decryption method.
\end{itemize}&
\begin{itemize}
\item Secure and performs effectively and efficiently.
\item Ensures secure communication from untrusted devices on the Fog network.
\item Achieved authentication, access control, verifiability and confidentiality.
\end{itemize}

\\ \midrule
Popa et al.~\cite{popa2010cloudpolice} & 
\begin{itemize}
\item A distributed multi-tenancy approach access control.
\item Access control only suits in infrastructure levels - as physical hosts and hypervisors.

\end{itemize}&
\begin{itemize}
\item Simpler, scalable and robust techniques.
\item Requires extra processing power.
\end{itemize}

\\ \midrule
Fan et al.~\cite{fan2017secure} & 
\begin{itemize}
\item CP-ABE based multi-authority data access control scheme in Fog-cloud computing systems.
\item Outsourced encryption and decryption computations.
\end{itemize}&
\begin{itemize}
\item User and attribute revocation can be performed efficiently.
\item Secure and highly efficient scheme.
\end{itemize}

\\ \midrule
Xiao et al.~\cite{xiao2017hybrid} & 
\begin{itemize}
\item A hybrid and fine-grained access control solution.
\item Most of the decryption process can be outsourced.
\item Secure and suitable in the Fog computing environment.
\item Perfectly applicable for resource-constrained IoT devices and applications.
\end{itemize}&
\begin{itemize}

\item Efficiency of data access is improved.
\item Key management cost is greatly reduced.
\item The limitation and drawbacks of this method is it can be applied only in centralized architecture.
\end{itemize}

\\ \midrule
Yu et al.~\cite{yu2018towards} & 
\begin{itemize}
\item Fine-grained access control and privacy is provided for Fog computing.
\item Can also guarantee security across side channel attacks.
\item leakage-resilient functional encryptions framework have been developed.
\end{itemize}&
\begin{itemize}
\item Highly secured and fine-grained access control.
\item Fully secure leakage-resilient functional encryption schemes have been presented.
\end{itemize}

\\ \midrule
Zaghdoudi et al.~\cite{zaghdoudi2016generic} & 
\begin{itemize}
\item Access control mechanisms proposed for Fog computing and ad-hoc MCC.
\item Focused on measuring the system overhead with different metrics.
\item A different size of networks, different hash function, and a variable responsible nodes percentage such metrics considered.
 \end{itemize}&
\begin{itemize}
\item A generic access control solution with features robust and scalable.
\item Take overhead with the increase of nodes in the network.
\end{itemize}

\\
        \bottomrule
    \end{tabular}
\end{table*}

\subsection{Malicious Attacks and Threats in Fog Computing}
Due to the isolated deployment of Fog nodes in some places, it fails to protect countermeasures and surveillances. As a result, it is very easy for intruders or malicious attackers to compromise the Fog networks through several malicious attacks~\cite{hu2017survey}. For example, a malicious user can compromise a Fog node with its own generated trust values, smart meter, smart grid, traffic system or spoof IP addresses~\cite{stojmenovic2014Fog} to ruin sensitive information. In this segment, we will give an overview of these potential threats and attacks issues.

\subsubsection{Potential Threats}

\begin{itemize}
\item \textbf{Rogue Fog Node:}
 Rouge Fog node is a one type of Fog device in Fog computing environment which presents itself as a legitimate node and persuades end users to connect with it. It may happen in such a scenario, when a Fog administrator instantiates an insider attack, to identify the rogue Fog node or legitimate Fog node. Stojmenovic et al.~\cite{stojmenovic2014Fog} have proven that the data can be tampered by a man-in-the-middle attack, with updated or collected the data  either in the Fog layer or cloud layer. There is also the possibility to launch additional attacks. So, in the context of privacy and security, the presence of a rogue Fog node will be a potential threat in the Fog environment. It is not easy to detect a rogue Fog node in Fog computing for various reasons. One of the main reasons is the diversified trust computing mechanism which brings about perplexed trust situations. On the other hand, we know that Fog computing is dynamic in nature, and consists of numerous devices which leads to creating, deleting, and revoking simultaneously. Therefore, for these various instances, it is difficult to manage the blacklisted nodes. The authors Han et al.~\cite{han2009measurement}~\cite{han2011timing} have demonstrated measurement-based models which permit a client to escape connecting to rouge access points (AP). Ma et al.~\cite{ma2008hybrid} introduced a framework to identify the existence of rogue APs in wireless networks. Detecting a rogue Fog node in an IoT network is cumbersome because of the network complexity across different scenarios~\cite{alrawais2017Fog}. Nevertheless, by using trust measurement-based models in the IoT network, it helps to detect rogue nodes. Although this method is not adequate, it can be considered for limited security protection.

\end{itemize}

\begin{itemize}
\item \textbf{Fault Tolerance:} Fog computing is an emerging distributed computing platform which consists of a huge collection of numerous devices which is widely geo-distributed and heterogeneous. Therefore, there might be high chance of failure of devices, as compared to cloud computing. Fog computing is dynamic in nature, whereby the Fog nodes or IoT devices connects or disconnects to a Fog layer over and over. Because of this behavior, there might be a chance to bring about unexpected faults and failures in the Fog environment. Therefore, in these circumstances, the Fog computing platform should provide all the necessary services without interruption if there is a failure occur in individual Fog devices, networks, applications, and services platforms~\cite{dastjerdi2016Fog}. Because Fog applications should be capable to instantly turn to other available nodes via some inbuilt mechanism if the services in an area become unusual. To mitigate these issues, standards should be applied. Stream Control Transmission Protocol (SCTP) is such example that can deal with such events and packet reliability in wireless sensor networks~\cite{madsen2013reliability}.

In general, fault tolerance ensures the availability of devices or applications in the event of a failure to provide uninterrupted services. Nevertheless, on the basis of what service is being used, fault tolerance will change according to one's role and management privileges. In the cloud computing environment, fault tolerance is handled by applying three techniques - proactive, reactive and adaptive~\cite{patra2013fault}.

\textbf{Proactive fault tolerance policies} refer to an escape rescue from faulty components by anticipating and replacing the failed components before it takes place.\\
\textbf{Reactive fault tolerance policies} refer to the decrease in the influence of faulty components when the failure occurs. In Adaptive fault tolerance, where the procedure is carried out according to the situation automatically.

There are numerous fault tolerance techniques which are often used in computing~\cite{latchoumy2011survey}~\cite{lussier2005fault}~\cite{bala2012fault} such as Replication, Job Migration, checkpoint, self-healing, Rescue workflow, Safety-bag checks, Task Resubmission, Software Rejuvenation, Masking, Preemptive Migration, and Resource Co-allocation. Nevertheless, in this paper, fault tolerance is mostly discussed based on the cloud computing environment as Fog computing is a new computing paradigm. In recent research works~\cite{wu2014model}~\cite{latiff2017checkpointed}~\cite{jiang2017fault}~\cite{liu2017framework}~\cite{sharma2016reliability}, the context of cloud computing in such a scenario was discussed. Therefore, fault tolerance in Fog computing is still a research task. In order to provide a reliable and robust Fog computing environment, failure handling of services should be effectively considered.
\end{itemize}

\subsubsection{Malicious Attacks}
Fog computing comprises various IoT or edge devices and collects the data from these devices by accomplishing latency conscious processes. Identifying malicious nodes is a complex task in the Fog environment~\cite{sandhu2017identification}. As we know, Fog computing is a miniature of cloud computing, as such, almost all types of malicious attacks, which affected a cloud environment can also affect Fog computing. For Example DDoS (Distributed Denial of Service), MITM, sniffing, side channel attacks, DoS (Denial of Service), malware injection, and authentication attacks attack are few of them. Therefore, in these circumstances, without an appropriate prevention mechanism, it can severely damage the competency of the Fog system or network. In this portion, we are going to expose a few malicious attacks which might occur frequently and affect the Fog environment.

\begin{itemize}
\item \textbf{Attacks from malicious Fog nodes and edge devices:} 
As Fog nodes are compromised easily by any malicious attacker, it is a very serious and potential threat for the Fog network environment. The authors~\cite{lee2015security} mention various unique security threats in their research, which might occur in the IoT and Fog environments. For delivering services to the users, the received data from the IoT devices will be processed by Fog nodes. If some Fog nodes are compromised by any intruders, it is a problematic task to ensure the security of the data. One possible solution would be, by establishing trust between Fog nodes themselves. In this case, an authentication mechanism is mandatory for ensuring secure, trusted communication. Therefore, Fog nodes cannot manage each other, so that it needs to trust only the cloud for  authenticity. Sequentially, after being authenticated by the cloud, it should be placed in a Fog environment to process heavy data. However, they are not able to give a suitable solution for this attack. 
Li et al.~\cite{li2017non}, carried out research and presented a solution.

It is vital to identify malicious Fog devices in Fog computing. Due to the lack of resource and edge devices, it is difficult to deploy proper authorization mechanisms between Fog nodes and edge devices. So, it is hard to prevent all attacks completely because of granting a few privileges and processing of the data. Sohal et al.~\cite{sohal2018cybersecurity} tried to solve the problem by using intrusion detection and virtual honeypot devices by introducing a Markov chain based framework.
\end{itemize}

\begin{itemize}
\item \textbf{Man-in-the-Middle (MITM) Attack:}
All data traffic passing through is protected through secure transmission channels between Fog nodes and edge devices in Fog computing. During this communication process, a user's data will be snooped or impersonated by an external malicious attacker prior to performing a global concealing process in the Fog node. Such a scenario correlates with the MITM attack. In a MITM attack, a perpetrator secretly relays and manipulates the data during communication between two parties. Hence, MITM is a potential attack method which can be used as a typical attack in Fog computing. In Fog computing, an attacker can carry out sniffing or disrupt the packets between Fog devices. As mentioned earlier, in Fog computing, all devices are resource constrained. By having this problem, it is becomes a challenging task to deploy secure communication protocols and encryption-decryption methods amongst Fog nodes and IoT devices~\cite{stojmenovic2016overview}. Stojmenovic et al.~\cite{stojmenovic2016overview} proposed an authentication method which can possibly avoid MITM attack. To mitigate MITM attacks, the anomaly detection is hardly applicable in Fog computing because these methods were being used in traditional cloud computing. Therefore, to mitigate MITM attacks in Fog computing, a compatible solution still offers a challenge, which can be considered for further research.
\end{itemize}

\begin{itemize}
\item \textbf{Distributed Denial of Service Attack (DDoS):}
In the modern epoch, Distributed Denial of Service or (DDoS) is one of the most renowned and challenging threats for cyberspace and other online services. As Fog nodes are made up of limited resources, it is troublesome to manage a huge amount of requests simultaneously. When a malicious attacker or intruder initiates a bunch of inappropriate service requests towards the targeted device, or tries to spoof multiple devices concurrently using the IP addresses, the Fog node will be occupied for a longer span of time. Therefore, all the legitimate services of Fog devices will be inaccessible for legitimate users. As opposed to, Fog nodes which go on to compromise themselves and get used for generating DDoS attacks. A different plane of the Fog environment can be affected by this kind of attack. Recently, malicious attackers have been able to compromise online home-automated smart devices to execute a DDoS attack against popular online websites such as Twitter, Paypal and Reddit. After these attacks, all of these websites were severely affected. Hackers have been trying to use internet-connected home automated equipment, such as Closed Circuit Television (CCTV) cameras, printers, refrigerator, etc. to perform DDoS attacks on popular websites, such as Twitter, Spotify, PayPal, SoundCloud and Reddit~\cite{bbc-cattack2016}~\cite{bbc-wattack2016}. In accordance with the Fog network system, all smart objects which are connected consists of more computational power and they have the ability to perform various tasks concurrently. As compared to traditional DDoS attacks, in Fog computing, various Fog devices apply DDoS attacks which will become much more severe. Therefore, it is not possible to mitigate a DDoS attack completely in the Fog computing environment. At the present moment, we can only monitor them. Under these circumstances, current DDoS issues may need new thinking and further research which will classify DDoS issue much more precisely in the context of the Fog computing environment.

\end{itemize}

\begin{itemize}
\item \textbf{Malicious Insider Data Theft Attack: }
According to the three-plane architecture of Fog computing, cloud computing is correlated to Fog computing. Hence, we should be conscious of all the malicious attacks which occur in cloud computing frequently. One severe attack in cloud computing could be a malicious insider attack for data theft purposes. On common terms, the end users will have to trust the cloud service provider despite being aware of this threat. It happens due to the deficiency of cloud service provider's authentication, authorization, and audit controls which allows attacks to spread out across the cloud system. In this regard, a few incidents have occurred which compromised corporate data, for example, Twitter's personal hacking ~\cite{arrington2009our},~\cite{takahashi2010french} as well as the account hacking incident of U.S. President Barack Obama~\cite{allen2010obama} which was exposed as a  malicious intent to steal a user's credentials. The authors Rocha et al.~\cite{rocha2011lucy} revealed that a malicious insider can gain access to the user's data easily in a cloud computing system. The attackers carry out their attacks which are generated from within cloud service providers. Therefore, the end user is not able to detect unauthorized access. There are diversified approaches which would be useful in order to secure data from faulty implementation, misconfigured service bugs in code by using encryption and access control to restrict them as well as to give protection from sophisticated attacks~\cite{pepitone2011dropbox}. Another solution could be user behavior profiling, where the system keeps track of the amount of user data access and the duration of data use. Hence, the system can identify anomalous activities of end users, which can be used to detect malicious attacks. In this case, the authors Stolfo et al.~\cite{stolfo2012Fog} have proposed a new approach to assure the security of cloud computing by using user behavior profiling and decoy technology. There might still be few issues~\cite{mukherjee2017security} which arise, on how to deploy the decoy in Fog networks and how to develop an on-demand decoy information to reduce the portion of stolen data from being lost.
\end{itemize}

\begin{itemize}
\item \textbf{Physical Attacks: }
In traditional data centers, physical security is being provided by on site security staff. On the other hand, by applying complex measures e.g. card punch, thumb impression, and retina scanning, physical access control can be deployed much more convincingly. So, these issues are related to certification and audits to derive the necessary physical security measures which are required to meet the set standards. Basically, Fog nodes are widely distributed across various environments. Due to point, it is impossible to implement traditional physical security measures in the Fog computing environment. For example, physical security measures can be applicable to place the edge box at the top of the streetlight's pole, which should be hidden from eye level as well as being surrounded with a fire-resistant coating to keep it safe from vandalism. There is a lower probability of physical attacks at the software level which enables the scope of theoretical attacks. 


\end{itemize}
In accordance with the study above, and based on different issues regarding threats and attacks related to the Fog, it can be summarized in Table \ref{sumAttack}. The focus of this study is to address auditing issues to secure the Fog computing environment. The following section discusses security auditing issues in Fog.

\begin{table*}[htbp]
	\centering
    \small
	\caption{The summary of threats and attacks issues in Fog environment from major survey papers}
	\label{sumAttack}
	\begin{tabular}{L{2.5cm}C{7cm}C{6.5cm}} 
		\toprule
		\textbf{Reference Paper} & \textbf{Highlights/Objectives} & \textbf{Achievement and Limitation} \\ \midrule

Stojmenovic et al.~\cite{stojmenovic2016overview} & 
\begin{itemize}
\item Managed to conduct a MITM attack.
\item This attack is very stealthy and dangerous.
\end{itemize}&

\begin{itemize}
\item An authentication scheme has been proposed to mitigate such attacks.
\item Encrypted communication method may not work always to protect from this kind of attacks.
\item On the other hand, complex encryption and decryption techniques are not always compatible due to resource limitation.
\end{itemize}
\\ \midrule

Wang et al.~\cite{wang2018Fog} & 
\begin{itemize}
\item Fog based storage technology to mitigate the cyber threat in the cloud.
\item Data stored separately in the Fog server as well as in the cloud storage.
\end{itemize}&
\begin{itemize}
\item Ensure the integrity, confidentiality, and availability of data.
\item Attackers unable to get any information about data by using data fragment.
\item Can protect the confidentiality of the user's data better than traditional ways.
\item This approach is safe and feasible for cloud storage.
\end{itemize}

\\ \midrule

Homayoun et al.~\cite{homayoun2019drthis} & 
\begin{itemize}
\item Fully automated and Fog node ransomware detection techniques for the Fog layer.
\item Deep learning techniques can be applied.
\end{itemize}&
\begin{itemize}
\item Detect and identify the ransomware within very short time execution of an application.
\end{itemize}
\\ \midrule

Han et al.~\cite{han2009measurement, han2011timing} & 
\begin{itemize}
\item The presence of fake Fog nodes or rogue Fog nodes is a serious threat to the Fog network.
\end{itemize}&
\begin{itemize}
\item A practical, timing based method for the end users to avoid connecting to rogue Access Point.
\end{itemize}
\\ \midrule

Stolfo et al.~\cite{stolfo2012Fog} & 
\begin{itemize}
\item Decoy technology and user behavior profiling have been used for disguise detection.
\end{itemize}&
\begin{itemize}
\item Mitigating insider data theft attacks.
\item Securing personal and business data.
\end{itemize}

\\ \midrule

Sandhu et al.~\cite{sandhu2017identification} & 
\begin{itemize}
\item A framework which uses three technologies such as an IDS, a Markov model, and a virtual honeypot device (VHD). 
\item Edge device classification depends on level of damage and frequency of attacks.
\end{itemize}&
\begin{itemize}
\item Proposed system is able to identify malicious Fog nodes in Fog.
\item Successfully identify the malicious devices and also decreases IDS false alarm rates of IDS.
\end{itemize}

\\ \midrule

Hosseinpour et al.~\cite{hosseinpour2016intrusion} & 
\begin{itemize}
\item Lightweight and distributed IDS system based on an Artificial Immune System (AIS).
\item Three-layered structure that includes the Fog, cloud, and edge layers.
\end{itemize}&
\begin{itemize}
\item Smart data approach has been used to build a lightweight and efficient IDS for the Fog platform.
\item Can detect silent attacks such as botnet attacks in IoT-based systems.
\end{itemize}
\\ \midrule

Alharbi et al.~\cite{alharbi2018focus} & 
\begin{itemize}
\item Security system based on Fog that defends the IoT system from malware attacks.
\item Proposed challenge-response authentication to protect IoT systems from further from DDoS attacks.
\end{itemize}&
\begin{itemize}
\item Able to filter malicious attacks effectively while response latency is very low and network bandwidth consumption is low.
\end{itemize}

\\
        \bottomrule
	\end{tabular}
\end{table*}

\subsection{Security Auditing in Fog}
In the traditional computing environment, it is often essential for technology experts to perform various security tasks such as examining security configurations, regulating potential vulnerabilities and constructing new security configurations with respect to every organization's own security policies~\cite{fu2017npp}. On the other hand, it is getting much harder when new computing paradigms like Fog computing are considered. Traditionally, organizations can enforce their access control policies according to its employee's roles and responsibilities, which is actually a challenging task for most administrators. Therefore, this challenge will be much more difficult in a Fog computing environment where security policies can be deployed across a huge number of devices residing at the edges of the Fog network. Security administrators need adequate knowledge to accomplish multifarious administrative tasks. Therefore, in this section, we discuss the various issues of Fog computing security auditing.

\begin{figure}[!t]
    \centering
    \includegraphics[width=3in]{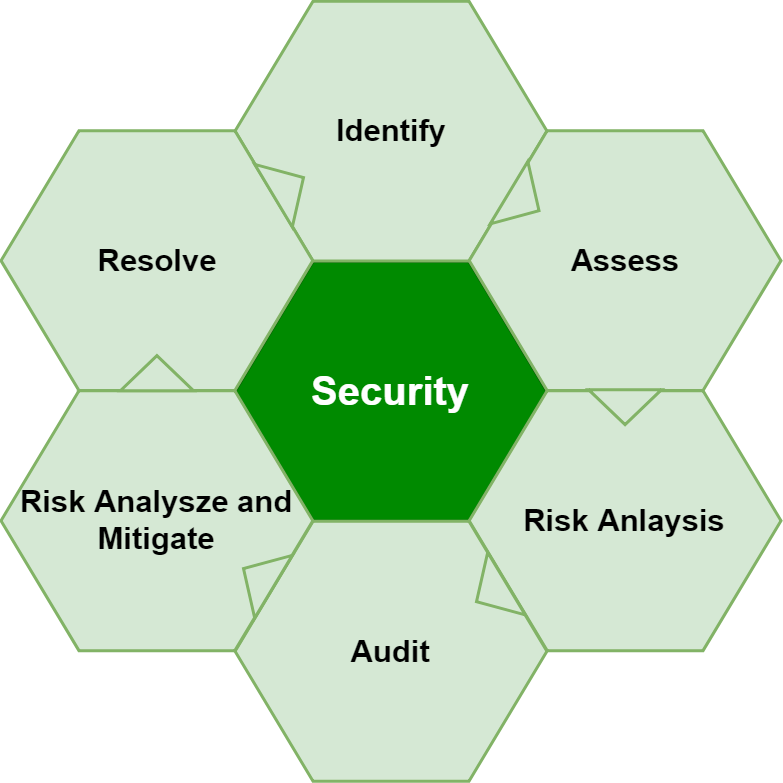}
    \caption{Life cycle of Security.}
    \label{fig_aud}
\end{figure}

\begin{figure}[!t]
	\centering
	\includegraphics[width=3in]{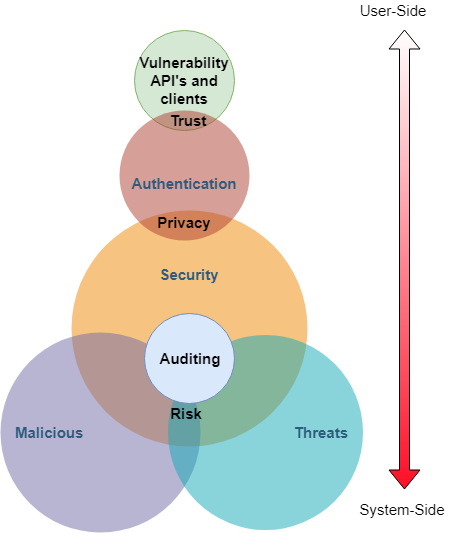}
	\caption{Risk level from user to system}
	\label{fig_Fogrisklevel}
\end{figure}

\textbf{Why is security auditing important for Fog?}\\
Fog computing is the latest computing paradigm in the modern computing world. The life cycle of security is shown in Fig. \ref{fig_aud}. The risk level from user to system is shown in Fig. \ref{fig_Fogrisklevel}. In spite of its substantial growth, there still remains lots of barriers for much more widespread adopting of Fog computing services due to security issues. Lack of auditability is a primary security concern in the Fog computing environment.


In the following section, we discuss several key aspects of Fog security auditing.

\textbf{Why is traditional security auditing not enough for the Fog?} \\ Fog computing has come up with numerous features and it is strongly dynamic in nature. All communication processes, data transmission, data analysis, user authentication, and resource management can be automated and dynamic with real-time operation. According to the nature of Fog computing, its security auditing process would be dynamic and within a real-time process. However, the existing traditional security auditing standards and the manner of auditing is very manual, where a technology specialist team or group of individuals perform their auditing processes using their traditional auditing standard. The traditional approach is only applicable within a small environment or with limited resource. However, it is a problematic approach because this approach provides only limited support to make an evaluation and the quality of the audit heavily depends on auditor's knowledge and experience. In such cases, several difficulties can be anticipated.

    \begin{enumerate}
    \item Security auditing expert's knowledge can be inadequate or inappropriate.
    \item To correctly configure out the Fog system's security, many organizations or users, find it cumbersome because of the extensive expenditure to hire security professionals.
    \end{enumerate}
    
    Therefore, a software-based automated auditing system, which can perform on a real-time basis, would be the best suited solution for the Fog computing environment.
    
    \textbf{How does Fog security auditing help to mitigate security breaches and privacy concerns?} \\
   Fog computing provides several security and privacy concerns for the cloud and traditional computing as well as its own security flaws. In the Fog environment, there are extensive amounts of devices, applications and resources which exist simultaneously and communicate with each other within a geographically distributed environment. Therefore, there exists a high opportunity for rapid security and privacy vulnerabilities. There are many security demonstrations which exists for traditional or cloud computing, but these demonstrations are not predominantly well-suited with respect to Fog computing. With the help of auditing Fog security configurations, we can mitigate these security issues as well as privacy-related issues for Fog nodes or Fog computing devices. Auditing security measures are a way of examining for infringement which potentially exposes the vulnerability of a system. 
   
    
   So, when one focuses on Fog based auditing, there is a need to see these concern as core to the overall approach:
    \begin{itemize}
    
        \item To minimize or mitigate risks introduced by Fog
        \item To identify new threats and defend them
        \item To evaluate the efficiency of security controls related to Fog
        \item To continuously improve policies, processes, procedures and tools
        \item To perform knowledge based dynamic periodic auditing processes
    \end{itemize}

\subsubsection{Criteria and Current Solutions}

Parkinson et al.~\cite{Parkinson:2017:SAF:3018896.3056808} proposed a novel Graph-based Security Anomaly Detection (Graph- BAD) approach that translates the object-based security configurations into a graph model. Another technique which was developed can identify vulnerabilities autonomously and perform security auditing of large systems without the need for expert knowledge.

Bleikertz et al.~\cite{bleikertz2010security} proposed an algorithm to audit the configuration network's security and the policies of the multi-tier cloud architecture using Amazon's EC2 public cloud.

Wang et al.~\cite{wang2013privacy} proposed an auditing system for data storage security by implementing a privacy-preserving auditing protocol using homomorphic authentication and random mask techniques for the preservation of privacy against TPA. It can audit without requiring to have the knowledge of the user's data contents. A batch auditing protocol was also introduced in this study, which can be used to complete multiple auditing tasks across different users at the same time via TPA. A public auditing system contains four algorithms such as, KeyGen, SigGen, GenProof, and VerifyProof. KeyGen is run by the user to set up the scheme, and to generate the required verification metadata, of which Siggen is used. GenProof is executed by the Cloud Server to provide proof of the data storage's correctness. VerifyProof is run by TPA to audit the proof from Cloud Server.

Cong et al.~\cite{wang2010toward} recommended a set of characteristics for public auditing systems with the aim to focus on data storage security in public cloud.

Shah et al.~\cite{shah2008privacy} proposed several public auditing protocols which helped not only to check data integrity from the service provider, but also fraudulent customers. Privacy preservation is achieved through zero-knowledge, and by concealing data contents from the auditor. Yang et al.~\cite{yang2012data} reviewed several current works on data storage security auditing service in cloud computing. Mohammed et al.~\cite{mohammed2018secure} proposed a secure protocol by a Third Party Auditor (TPA) that ensures the data integrity in Fog computing. The main drawback of this method is that the user has to depend on a third party. There should be trust between the Third Party Auditor(TPA) and user.

\subsubsection{Existing Security Auditing Standards and Frameworks}
Implementing security governance and auditing frameworks may support organizations to conduct and manage their own security risk levels. Various organizations or technology groups have created renowned frameworks and recommendations based on the traditional computing or cloud computing standards \cite{spremic2011standards, ryoo2014cloud} which are globally used. Therefore, the most popular and renowned security audit standards and frameworks are as follows:

\begin{itemize}
\item \textbf{Service Organization Control (SOC) 2:} which is considered for auditing outsourced services sponsored by the American Institute of CPAs
\item \textbf{ISO 27000 standards - ISO 27001:2005 and ISO 27002:2005 :} Traditional security audits sponsored by ISO
\item \textbf{CobiT (Control Objectives of Information and related Technology):} sponsored and introduced by ISACA(Information System Audit and Control Association, www.isaca.org) and ITGI (IT Governance Institute, www.itgi.org). It is the most renowned and extensively accepted information technology governance framework.
\item \textbf{NIST (www.nist.org) 800-53 revision 4:} Federal government audit sponsored by the National Institute of Standards and Technology (NIST)
\item \textbf{Cloud Security Alliance (CSA):} Cloud-speciﬁc audit which is presented to cloud security auditing terms sponsored by CSA
\item \textbf{Payment Card Industry (PCI), Data Security Standard (DSS): }PCI Qualiﬁed Security Assessor cloud supplement which is sponsored by PCI DSS
\item Basel II, ITIL,  SANS(www.sans.org), (ISC)2 framework (www.isc2.org), etc organization which can audit and manage the levels of IT security risks.
\end{itemize}

To be effective, the above-mentioned security audit standards must confirm to a vast number of security concerns in the traditional computing or cloud computing paradigm. However, using these traditional auditing standards and frameworks in the Fog computing environment will not be well suited because all of these auditing standards and frameworks which are manual approaches. They can only provide limited support to make an evaluation and the audit's quality heavily depends on an auditor's experiences and knowledge which could be problematic, whereas the Fog environment is mostly dynamic and distributed across a large scale geographically. Therefore, software based automated auditing standards and frameworks which can perform real-time approaches would be best suited for the Fog computing environment.

The principal necessity to introduce cooperative context aware tools is extensively approved, and actions are being taken at the state level. Several studies have suggested how software tools can be used to extract meaningful knowledge to aid security configurations, auditing, and digital investigations \cite{franke2014cyber}. Therefore, such tools are context-dependent, in that their functionality is conducted to identify threats that are expected. The only limitation of these tools is that each one requires different knowledge and skills to translate their output to obtain an understanding of why this extracted knowledge is significant \cite{garfinkel2010digital}. Security auditing can be performed in an automated fashion by using Blockchain technology. The next section discusses Blockchain technology and what has been done so far in Fog using Blockchain technology.







\section{Blocakchain Technology in Fog}
\begin{figure}[!t]
	\centering
	\includegraphics[width=3in]{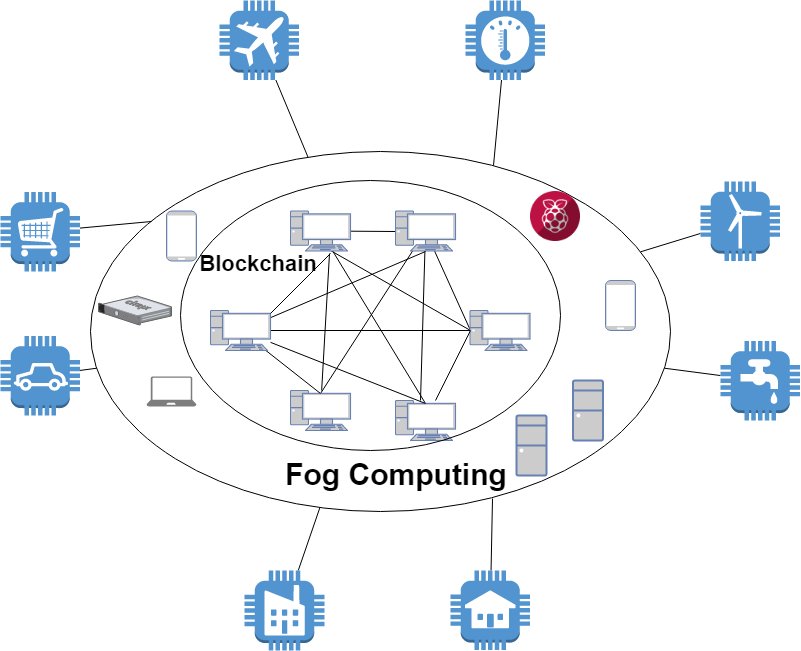}
	\caption{Fog with Blockchain}
	\label{fig_Fogblo}
\end{figure}

The Blockchain is more than a database technology. Theoretically, a Blockchain is a ledger of the distributed database that can be programmed continuously to record a list of data. Blockchain is probably Bitcoin's major innovation foundation for a new decentralized and distributed system. Recently, Blockchain technology has been implemented across many real-time systems  \cite{zyskind2015decentralizing}. Blockchain is an evolving technology to build a secure, scalable and openly coordinated platform globally, which is not limited to currency or financial systems. Fog with Blockchain is shown in Fig. \ref{fig_Fogblo}.   

\subsection{Security Features of Blockchain Technology}
Blockchain technology has its own strong security because there is no possibility of shutting down the system. A well-known cryptocurrency - Bitcoin, was implemented using Blockchain technology. However, the financial system was still hacked, of which it has never been subjected to before. The main strength of Bitcoin is its use of the Blockchain network which is protected against attacks and threats by using multiple nodes which are committed to a single transaction by a consensus algorithm on this network. The transaction within Blockchain includes digital signatures. Currently, Blockchain uses the ECDSA public key algorithm to generate a digital signature. Blockchain prevents a single point of failure because it is a distributed system. It uses a hash function for block generation, of which currentlyit uses the SHA-256 hash function.

Some of the main features of Blockchain are as follows:
\begin{itemize}
    \item Increased Capacity
    \item Strong Security
    \item Immutability
    \item Faster Settlement
    \item Decentralized System
    \item Offers encryption and validation
    \item Virtually impossible to hack
    \item Can be private or public
    \item Minting
\end{itemize}

\subsection{Role of Blockchain to Improves Security in Fog}
 The Blockchain technology was introduced for the secured cryptocurrency application Bitcoin. A realization soon dawned amongst many researchers that it possesses great security features which can be utilized in many real-world distributed applications (e.g. Cloud, and Fog computing). Security has become a key stumbling block toward the widespread adoption or implementation of Fog. Therefore, security concerns in Fog computing can be improved using Blockchain technology
\footnote[1]{https://securitytraning.com/how-blockchain-can-improve-iot-security/}
\footnote[2]{https://businessinsights.bitdefender.com/blockchain-improve-internet-of-things-security}.
\footnote[3]{https://blogs.cisco.com/innovation/blockchain-and-Fog-made-for-each-other}
\begin{itemize}
    \item Mitigate single point of failure
    \item Highly encrypted network transactions
    \item Node status tracking capabilities
    \item Immutable Technology
\end{itemize}

Blockchain can mitigate various threats and attacks in Fog such as the man in the middle attack, DDoS attack, and data tampering
\footnote[4]{https://bdtechtalks.com/2017/01/11/how-blockchain-can-improve-cybersecurity/} \footnote[5]{https://cybersecurityventures.com/how-blockchain-can-be-used-to-improve-cybersecurity/} \footnote[6]{https://securitycurrent.com/four-ways-improve-security-blockchain/} \footnote[7]{https://www.esecurityplanet.com/network-security/blockchain-security.html}.

\subsection{Blockchain between Fog and Edge Environments}
Fog computing is a decentralized distribution system which aims to make cloud computing faster by creating data hubs or mini data processing centers which are hosted in smart devices. Basically, they accomplish a less demanding task and reduce the communication between the cloud and the end user. Fog allows performing resource-constraints and short-term analytics near to the edge of the network, whereas the cloud accomplishes resource-intensive and longer-term analytics.

Fog computing faces enormous challenges and there are constantly various issues which arise during its primary stages of development. For example, in a distributed computing environment it is a fact that how to protect its transactions and network resources with an evenly distributed security architecture is a challenge. It builds a kind of mesh network where every Fog node takes part based on their strength. Due to the distributed architecture of Fog computing, it is highly required when trust and security must be distributed. This is particularly significant where the Fog infrastructure, layers and Fog nodes are managed and owned by diversified entities.

However, a significant question arises in managing trust in a distributed and decentralized manner amongst participants that do not need mutual trust. Blockchain technology in reality is built for this kind of challenge. Blockchain consensus algorithms have a suitability issue with regards to Fog applications. For instance,
``Proof-of-Work'' (PoW) consensus needs a huge computing capacity in order to solve a complex mathematical puzzle, so Fog devices are unable to host this mechanism. But there are plenty of other protocols such as ``Proof of Stake'' (PoS) which is susceptible to running on Fog nodes with a similar capacity.



\subsection{Recent Works that Used Blockchain for Fog}
Tuli et al.~\cite{tuli2018Fogbus} developed a framework which was based on blockchain for the edge-Fog computing environment. This framework applied blockchain, encryption techniques and authentication which can perform secure operations across sensitive data. Although this framework is a lightweight and based on a cross-platform, it has a few limitations and drawbacks because it takes comparatively higher computational overhead to carry out large scale deployments.

Sharma et al.~\cite{sharma2018software} introduced a new and efficient distributed blockchain cloud model based on three emerging technologies: blockchain, Fog Computing and Software Defined Network (SDN). This model was presented to support high scalability, security, high availability, resiliency, real-time data delivery and low latency.

Jeong et al.~\cite{jeong2018security} proposed a blockchain based secure Fog computing system. Their system can defend against various attacks such as IP spoofing, Sybil attacks and single points of failure. This system used the Blockchain method to guarantee secure authentication and non-repudiation. It can also perform when a Fog node is down.

Samaniego et al.~\cite{samaniego2016using} investigated the idea of virtual software-defined IoT components known as virtual resources in combination with the use of blockchain technology.

Dorri et al.~\cite{dorri2016blockchain} introduced a secure, private and lightweight blockchain-based technology for the resource constraints related to IoT devices which can handle most security and privacy threats. It uses different kinds of blockchains based on the network hierarchy and uses distributed trust methods to assure a decentralized topology.

\subsection{Blockchain Oriented Startups in Fog and IoT Environments}
OpenFog Consortium is one of the most well-known Blockchain oriented startups in the Fog environment. The OpenFog consortium is in the process of building a composable and interoperable framework for Blockchain in the Fog distributed system. That implies that the various entities in the system that do not trust or are even known to each other still provide a meaningful consensus algorithm which is able to make decisions in a Fog oriented distributed system. The ”autonomy” which is one of the eight pillars of OpenFog, is supported by the Consortium's work.

Recently, there have been multiple Blockchain oriented startups which have joined the OpenFog Consortium \cite{bc2018helder} they are as follows:
\begin{itemize}
\item\textbf{iExec:} It Is the first Blockchain-Based Decentralized marketplace for Cloud Computing. It provides distributed applications that are secure, easily accessible and scalable to the services of computing resources for data-sets that are needed as well as the systems running on Blockchain (DApps). 

\item\textbf{KeyChain:} A new Global Blockchain-based data security infrastructure. It provides secure decentralized data authentication for the enterprise, finance environments, industries, and IoT.

\item\textbf{Aetherworks:} Brings original, high-quality technologies to the market and provides original software for distributed systems, including Fog computing and software-defined storages.

\item\textbf{Hyperchain:} Provides an enterprise-level Blockchain network-based solution for government agencies, supply chain, data trading, fraud prevention, and securities. It also supports enterprises to rapidly deploy, expand and configure Blockchain networks based on the Blockchain cloud platform.

\item\textbf{SONM:} Provides infrastructure and can run any decentralized application (Fog application) or host Blockchain-based services. It also provides Fog computing distributed cloud computing services such as IaaS and PaaS, which are secured by Blockchain.

\item\textbf{Xage:} The foremost Blockchain-protected security tool for the industrial IoT. Traditionally, more points of security vulnerability arise when there are more nodes and more connections. Moreover, the centralization technology prevents industrial systems working independently and in real time. Xage ensures that with the combinations of Blockchain and encryption that more nodes mean more security, not less. 
\end{itemize}

\section{Research Challenges and Future Research Direction}
In this section, we are going to present and highlight a few significant and considerable issues which are challenging tasks for Fog computing to cope with in cloud and edge environments. Finally, we provide a synopsis of probable research directions based on the existing research challenges.

\subsection{Trust Management}
Identification of trusted Fog nodes is a challenging task in the Fog platform. Usually, a Fog node is trusted or untrusted can be identified by its malicious behavior. But in this case, the malicious nature is not defined earlier-on for a Fog node. Therefore, it is significant to define and categorize all malicious characteristics in the Fog system. The Fog system can be susceptible to regulate if a Fog node is trusted or untrusted. Hence, it is mandatory to enhance trust and after all an exalted trust management model is highly required.

Another challenging research issue is, combining both distributed and centralized environments which is must and important in the context of cloud-Fog-IoT environments. Therefore, a centralized trust management is required for the IoT environment and it would be possible by using a Fog platform. Hence, it's still a research issue.

Moreover, trust management in Fog platform is entirely different compared to the cloud computing platform due to the distinctions of the cloud and Fog platform architecture and services offering mechanism. As mentioned earlier, Fog is widely distributed, on the other hand, cloud is centralized. In that case it is easier to deploy trust management in the cloud environment because the cloud platform has its own in-place security infrastructure, whereas the Fog platform is more open, and the in-place security mechanism is absent. As a result, the Fog is vulnerable to malicious attacks. In addition, trust in the cloud environment is unidirectional, whereas trust in the Fog environment would be bidirectional in nature. The Fog node and the IoT devices must maintain a trusted relationship between one another before their interaction, as it is highly required in the Fog platform. Hence, designing a bidirectional trust model in the context of Fog and the IoT platform is a challenging task as well.

\subsection{Privacy Assurance}
The Fog nodes hold sensitive or private information of users, as the Fog nodes are placed in the proximity of the end users. Therefore, it is a challenging issue to assure trusted communication and make a secure computing environment between the Fog and IoT devices. In such a case, we can consider encrypting the user sensitive data before sending it to the Fog nodes. It is not viewed as a proper technique in the context of IoT devices, since conventional encryption and decryption mechanisms need much computational power, whereas the IoT devices faces challenges to encrypt and decrypt the user's sensitive data due to the resource constraints of IoT devices.

In another context, a single Fog node can manage sensitive data which comes from different Fog users or across different applications. Therefore, there might be a chance to mix up different sources of data after the data aggregation step. In such a case, enforcing proper data encapsulation techniques at the Fog API or middleware level would be the solution. Hence, more research is needed.

Another challenging issue is to provide context-aware services in the Fog environment to the end user devices which are often involved in sharing sensitive resources such as location, as well as others personal information amongst other geographically connected devices. Therefore, in such a scenario, it is highly required to ensure data protection is present. Hence, providing the identity and location privacy in the Fog environment is a challenging task.

\subsection{Authentication}
It is an obvious fact that strong authentication and secure communication protocols in the Fog platform are missing. It is a rather alarming message for the research community. There has not been much research about the authentication mechanism in the area of Fog computing. Although, several researchers have already proposed several solutions which we described earlier in the taxonomy section. However, those solutions are still not able to cope with the Fog platform. Therefore, to design and develop a new authentication method for Fog computing, one must consider the following criteria and how that it can cope up with the Fog platform smoothly.

\begin{itemize}

    \item Authentication mechanisms must be compatible with the Fog user, end devices(IoT devices), application services and Fog Service providers on the cloud-Fog-IoT platform.
    \item Conventional authentication mechanisms are inefficient, and there is a necessity for a secure, environment-friendly, efficient, and scalable solution to cope up with extensive amount of IoT devices which has limited resource to facilitated scalability and efficiency.
    \item Security and performance are both highly required in terms of different contextual devices and applications.
    \item Must meet the dynamic behavior of the Fog environment, where Fog nodes dynamically leave and join frequently in the Fog network.
    \item Must ensure low complexity-based authentication in terms of scalability of the Fog network.
    \item Ensure smooth authentication and re-authentication methods in a dynamic manner.
    \item Design an efficient authentication method, of which a cryptographic lightweight encryption algorithm should be considered between the Fog system and the IoT devices that can easily cope with the low processing power of IoT devices.
    \item Authentication should be less costly, as well as provide high usability and in return should be user friendly.
\end{itemize}

\subsection{Access Control}
In terms of the authentication mechanisms, there has not been much  research work about access control methods in the Fog computing environment. However, plenty of work has been done in this field. Therefore, we still need to be able to accomplish an efficient design to draw the right kind of potential access and control model, with the intention to facilitate a secure platform within the heterogeneous devices in the Fog environment.

In the description section of access control, we mentioned a few access control models, describing their various features, characteristics and in the context of the Fog environment, we also highlighted numerous drawbacks and limitations. Many researchers have mentioned that Attribute Based Encryption(ABE) would be suitable as a method of owning access to control in the cloud, Fog and IoT environments. Because of the heterogeneous characteristics of the Fog system, the ABE method should be reconstructed in order to mitigate the major challenges (Latency, policy-management, fine-grained and enforced by the cryptographic method) amongst the Cloud-Fog-IoT computing environment users.

On the other hand, in the Fog system, data  originates, is encrypted and decrypted by miniaturized devices with low computational powers. In such a case, deploying access control mechanisms in that devices would be a burden and would need heavy computational powers to process the access control mechanism. Meanwhile, Fog devices are being placed near end devices. In addition, Fog devices are much more computationally powerful than end user IoT devices. Therefore, to overcome the limitations of IoT devices, an outsource capability lightweight ABE based access control would be compatible with the Fog environment. As opposed to, Fog computing, which is dynamic in nature, there are numerous devices which join and leave simultaneously in the Fog network., So, the access control policy and attributes of the users would be changed according to this dynamic characteristic. Therefore, it is highly required that ABE-based access control mechanisms must have the capability to assist in creating, updating, as well as revoking the attributes of the users. With ABE based access control, designing the revocation process would faces new challenges, and how Fog collaborates with the cloud environment during the revocation process would need to be part of further research. 

Therefore, to design a new access control method for the Fog platform, one must consider a few characteristics which are as follows:

\begin{itemize}
\item As we have mentioned earlier, Fog is a fully virtualized platform by nature and it provides diversified environments for the Fog network. In this case, there might be a chance in which a side-channeled-attack occurs due to the nature of sharing resources amongst untrusted tenants. Therefore, it is a significant concern
in terms of designing an access control method which must be capable to synthesize within the virtualized platform and multitenant environment efficiently, and securely.
\end{itemize}

\begin{itemize}
\item Access control should be secure and efficient for the Fog environment computing on the basis of multi-authority, as well as attribute-based, considering low computation with outsourcing capabilities as well as attributes have the means to control user revocation capability.
\end{itemize}

\begin{itemize}
\item An access control method should be lightweight and fine-grained due to the resource constraints suffered amongst IoT devices.
\end{itemize}

\begin{itemize}
\item An access control method must be capable to perform in both centralized and distributed architectural environment accordingly.
\end{itemize}

\subsection{Threats and Attacks} As we mentioned earlier, Fog computing faces various security and privacy issues. Due to the distributed nature and extensive amount of devices connected with it, often, there might be a chance for a threat or an attack to occur. In the description section, we have already highlighted several threats and attacks and their impact in the Fog environment. Detection, identification and mitigation of these threats and attacks would be a challenging task in terms of the dynamic Fog computing environment. However, in order to build a reliable and trustworthy Fog platform, there is a research gap and the lack of security solutions available to detect and identify these threats and attacks needs to be addressed. Based on our review across various threats and attacks, we have suggested the following issues which need to be addressed in the future to overcome these challenges:

\begin{itemize}

    \item Complex trust situations and insecure authentication and authorization systems.
    \item Dynamic behavior such as creating, deleting, joining and leaving of Fog nodes, or servers in the Fog layer.
    \item Detection of malicious nodes or rogue nodes is a challenging task because of the dynamic nature of leaving and joining by the Fog nodes.
    \item Implementing IDS in large-scale, geo-distributed with low-latency requirement with highly  mobile Fog computing systems is a complex task.
    \item Due to the distributed environment, hybrid detection techniques are required to identify malicious activities.
    \item Due to the resource constraints of the Fog devices, designing a high security and low cost threat and attack detection is the key problem in the Fog.
    \item Identification and mitigation threats and attacks from both the Fog node and Fog user at the same time is challenging.
\end{itemize}

\subsection{Security Auditing}
Audit rights provide a crucial risk mitigation tool regarding security issues related to the Fog. Auditing security configurations in the Fog platform is a complex task, as it is a gateway to the cloud platform and heavily relies on expert knowledge, which is required for understanding the different security configurations. However, these systems can be imperfect, and not user friendly for the home users and small companies.   

In this Section, we explore various unique challenges that isolate Fog security auditing from the traditional security auditing or cloud security auditing protocols. These challenges represent the significance of special provisions for Fog security auditing in current or evolving security auditing standards.

\textbf{Challenges:}
\begin{itemize}

    \item The Fog computing landscape is dynamic and consists of huge resources, where traditional data encryption or decryption needs heavy computational overhead.
    \item Without proper technological support, it is challenging to manage extensive amounts of different contextual data.
    \item To identify new security threats and defend against those threats is also a challenging task
    \item Fog computing brings easy accessibility to our work and personal lives, but with that accessibility comes new security risks and challenges
    \item Understanding the different contexts of the Fog computing environment is important. Different contexts with regards to the environment's security issues would bring about different. 
\end{itemize}

\textbf{Questions:}
\begin{itemize}
    \item How to encrypt or decrypt data and how to access that data simultaneously?
    \item How to perform auditing processes across different environment data contexts?
    \item Do you use the same matrix for the edge environment or cloud environment?
    \item Can your current risk assessment capture the risks correctly?
    \item How to perform and manage real-time processing and auditing at the same time?
\end{itemize}

In order to overcome the above-mentioned challenges and questions, it is highly required to develop an automatic method, which can be capable to recognize and identify security infringements as well as mitigate those security risks in Fog computing. Further research needs to be carried out by utilizing Blockchain technology to mitigate security issues in Fog. 

\subsection{Secure 5G Enable Fog Network}
In the near future, Fog devices will be connected through the 5G network. Connecting Fog devices with 5G network emerging new security challenges in mainly in the authentication. The traditional one-way or mutual authentication process is not useful due to the authentication process between the user and services \cite{Huawei5g}. In this case, a new hybrid authentication model is required. Using 5G, Fog will be useful to talk with things and devices. For example, in a smart home and smart city environment, one citizen needs an ambulance which will direct him to a specialized hospital near to the location of the user where a remote surgery can perform. Here, a hybrid security mechanism is required to secure the whole application environment since many parties are involved in this processing. Any emergency environment similar to this requires a strong and reliable authentication process. User privacy is also important in such 5G enable Fog computing environment. Because, user data may pass through the various untrusted, third-party devices, network equipment, and access networks. Hence, we need to explore more about hybrid authentication methods and privacy protection in 5G enable Fog network.

\section{Conclusion}
The main objective of this study is to review, investigate and analyze the issues of the Fog computing platform to recognize their probable security flaws. The obvious fact is that, there are numerous security issues that did not exists in the traditional cloud computing environment, of which need to be considered, as well as significant developments in the Fog environment. We fill the gap of the current literature by aggregating all security aspects of Fog computing paradigm. We have also investigated the main challenges, and tried to exhibit the motives as to why the security methods in the cloud platform cannot be employed directly in Fog computing when it comes to auditing. In this study, we have introduced a taxonomy, by considering numerous security issues and protection according to the Fog environment, as well as briefly introduced and discussed these issues retrospectively. In addition, we also discussed how blockchain could help to provide solutions to some of the data security concerns in the Fog environment. At the end, we highlighted several threats and attacks which might be occur frequently under the circumstances of the Fog computing network.

Interestingly the Fog is a new paradigm, which therefore requires mitigation of the associated security issues which are still challenging tasks. With regards to the system architecture of Fog computing, researchers need to do further future work and figure out the challenges with respect to security within the three tier architecture of the cloud-Fog-IoT computing system. As Fog computing is an extension of cloud computing, in this paper we only covered the security issues concepts related to Fog. We did not consider the security-related issues in the cloud.

In the future, we will be investigating and comparing and other similarly distributed environments and present these security issues and suitable solutions for the Fog.



%
%


%




\ifCLASSOPTIONcaptionsoff
  \newpage
\fi



\bibliographystyle{IEEEtran}
\bibliography{paper.bib}
%



%

\begin{IEEEbiography}[{\includegraphics[width=1in,height=1.25in,clip,keepaspectratio]{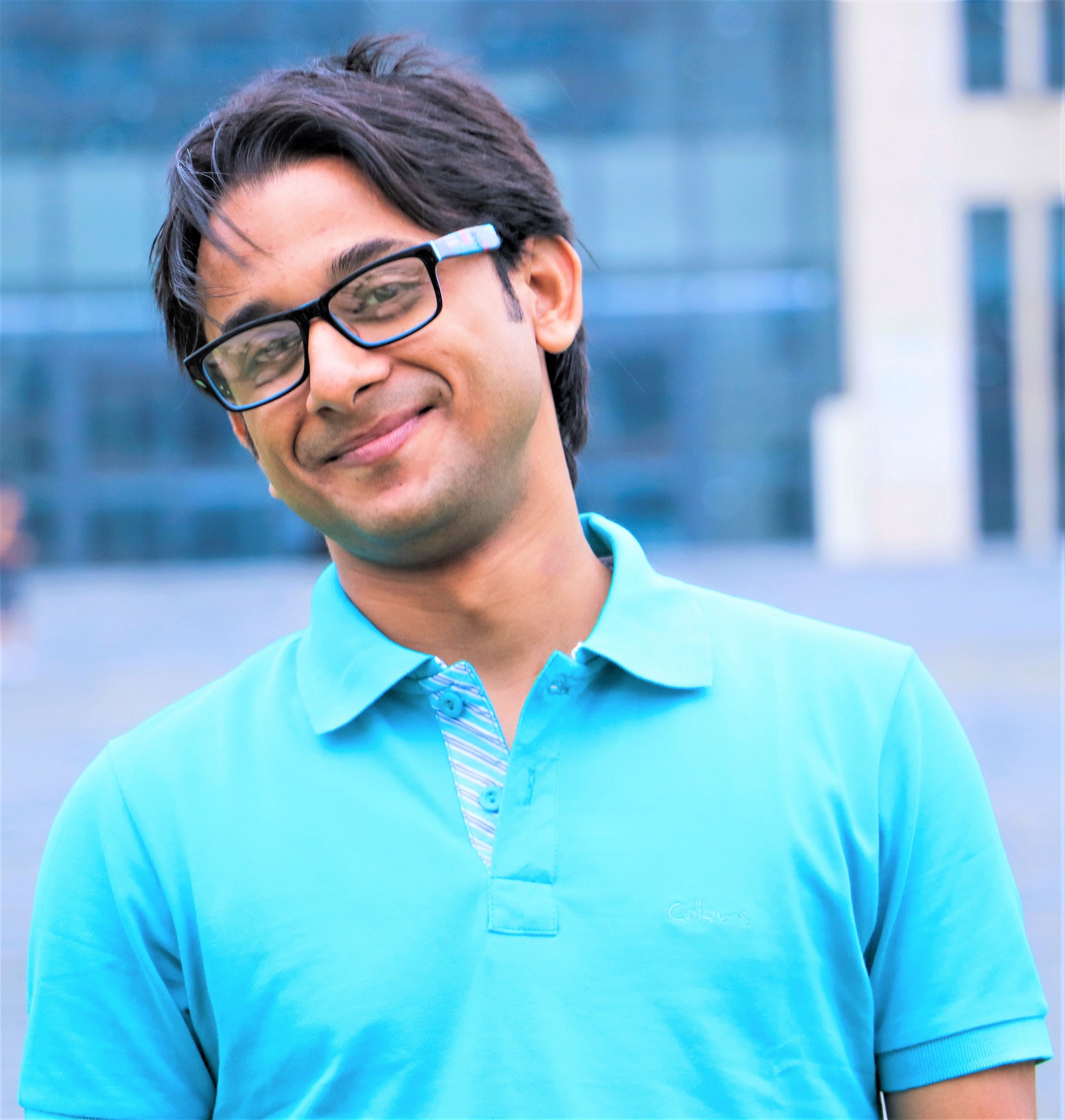}}]{Abdullah Al-Noman Patwary} is currently pursuing the Master's degree in the field of computer science and engineering from Nanjing University of Science and Technology. He is actually well-versed in most things network or information security related. His research interests include Fog Computing security, IoT security and Cloud Computing security. He received the bachelor's degree in computer science and engineering from State University of Bangladesh in 2014. Since 2014-2016, he has been working as a system administrator at Creative IT Ltd, Bangladesh. 
\end{IEEEbiography}

\begin{IEEEbiography}[{\includegraphics[width=1in,height=1.25in,clip,keepaspectratio]{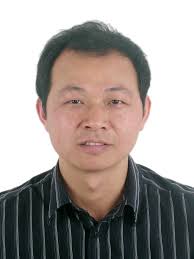}}]{Anmin Fu} is an associate professor and supervisor of Ph.D. students of Nanjing University of Science
and Technology, China. He received his B.S. degree in Communication Engineering from Lanzhou University of Technology, China, in 2005. He received his M.S. and Ph.D. degrees in Cryptography and Information Security from Xidian University in 2008 and 2011, respectively. His research interests include cloud computing security, wireless security
and applied cryptography.
\end{IEEEbiography}

\begin{IEEEbiography}[{\includegraphics[width=1in,height=1.25in,clip,keepaspectratio]{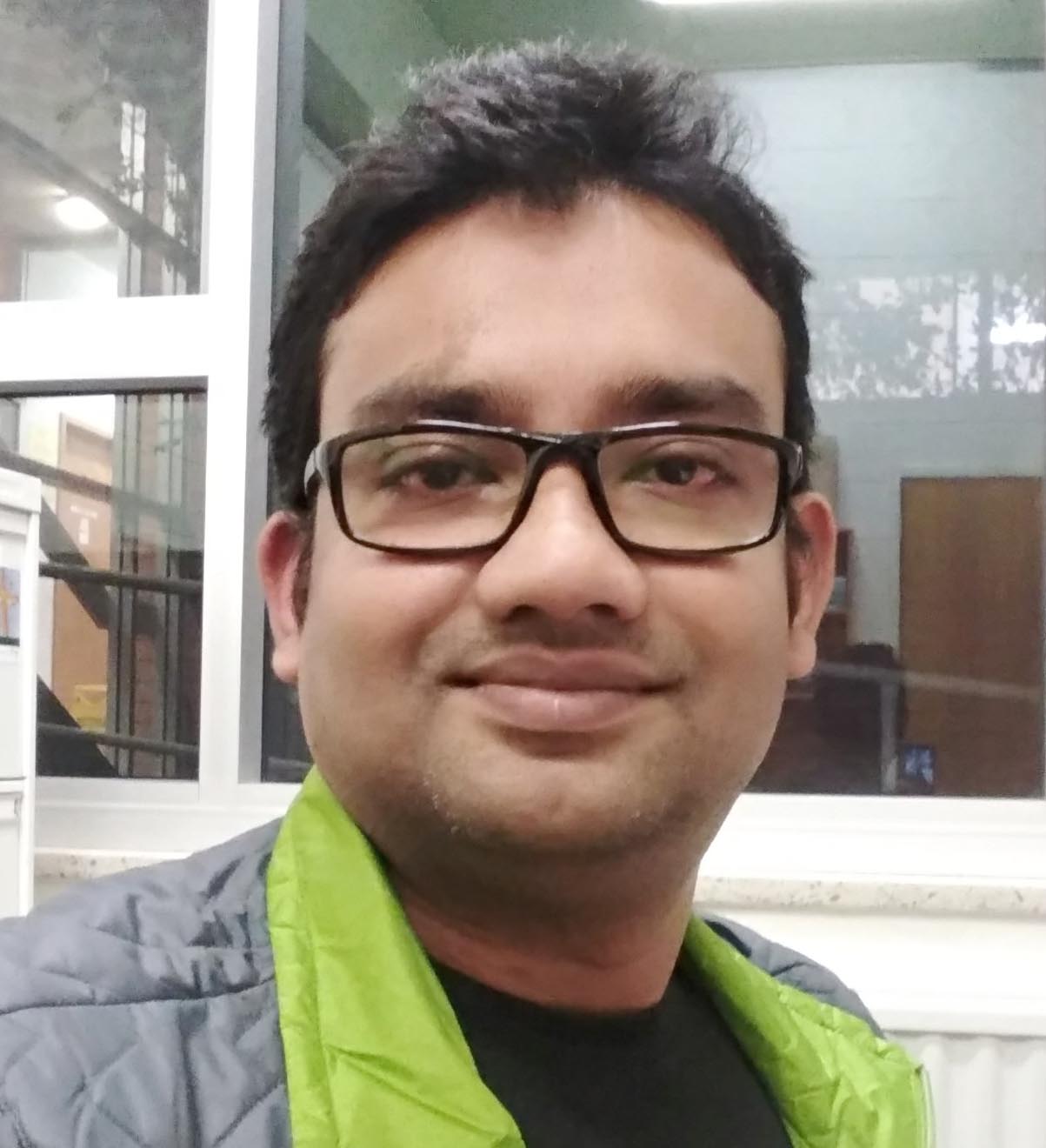}}]{Ranesh Kumar Naha} is currently pursuing his Ph.D. studies on reliable resource allocation and scheduling in Fog computing environment with the University of Tasmania. He has been awarded Tasmania Graduate Research Scholarship (TGRS) for supporting his studies. His research interests  include wired and wireless network, parallel and distributed computing, Cloud computing, Internet of Things (IoT), and Fog computing. He received his Master of Science (M.Sc.) degree from Department of Communication Technology and Network, Faculty of Computer Science and Information Technology, Universiti Putra Malaysia, in 2015. He received B.Sc. degree in Computer Science and Engineering from State University of Bangladesh in 2008.  During his master study he has been awarded Commonwealth Scholarship provided by Ministry of Higher Education, Malaysia. He served as Lecturer until 2011 in Daffodil Institute of IT, Bangladesh.
\end{IEEEbiography}

\begin{IEEEbiography}[{\includegraphics[width=1in,height=1.25in,clip,keepaspectratio]{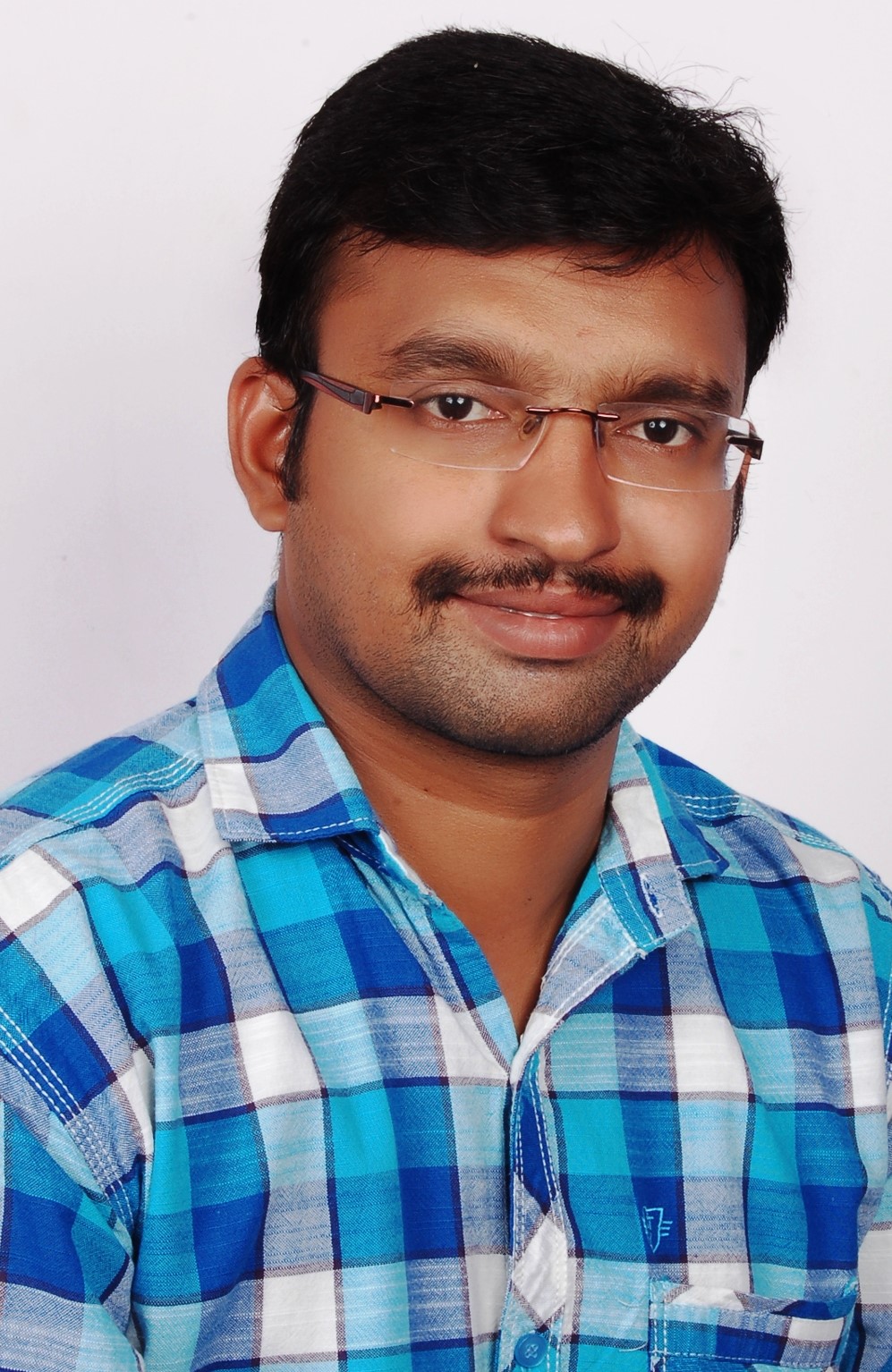}}]{Battula Sudheer Kumar}
received his Master of Technology degree in software
engineering in 2012. He is currently pursuing his Ph.D. studies on resource management in Fog computing environment with the University of
Tasmania. He has been awarded Tasmania Graduate Research Scholarship (TGRS) for supporting his studies. His research interests includes Fog computing, Distributed file systems, Cloud computing, Internet of Things (IoT), and Big Data.
\end{IEEEbiography}

\begin{IEEEbiography}[{\includegraphics[width=1in,height=1.25in,clip,keepaspectratio]{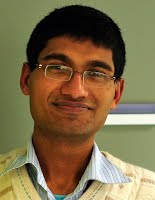}}]{Dr. Saurabh Garg} is currently a Lecturer with
the University of Tasmania, Australia. He is one
of the few Ph.D. students who completed in less
than three years from the University of Melbourne.
He has authored over 40 papers in highly cited
journals and conferences. During his Ph.D., he has
been received various special scholarships for his
Ph.D. candidature. His research interests include
resource management, scheduling, utility and grid
computing, Cloud computing, green computing,
wireless networks, and ad hoc networks.
\end{IEEEbiography}

\begin{IEEEbiography}[{\includegraphics[width=1in,height=1.25in,clip,keepaspectratio]{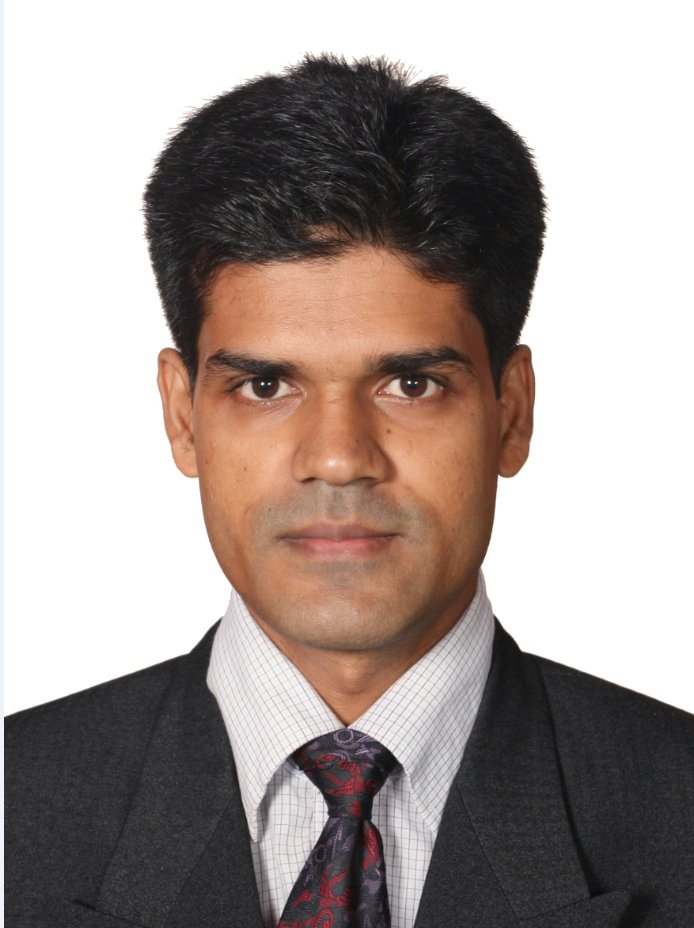}}]{Md Anwarul Kaium Patwary} completed his Master of Science in Computer Science from the Universiti Putra Malaysia. He is currently pursuing a PhD in Computer Engineering at the University of Tasmania. His research interests include dynamic graph partitioning, graph algorithms, load balancing, and distributed computing.  
\end{IEEEbiography}

\begin{IEEEbiography}[{\includegraphics[width=1in,height=1.25in,clip,keepaspectratio]{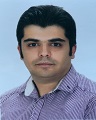}}]{Erfan Aghasian}received the B.Eng. degree in information technology from Qazvin Azad University and the M.Sc. degree in information technology management from the University Technology of Malaysia. He is currently pursuing the Ph.D. degree in information technology with the  University of Tasmania. His research interests include computer systems and network security, data security and data anonymisation.

\end{IEEEbiography}







\end{document}